\documentclass[preprint,11pt,authoryear]{elsarticle}
\usepackage[margin=1in]{geometry}
\usepackage{setspace}
\usepackage{amsmath,amssymb}
\usepackage{booktabs}
\usepackage{graphicx}
\usepackage{caption}
\usepackage{subcaption}
\usepackage{threeparttable}
\usepackage[table]{xcolor}
\usepackage{rotating}
\usepackage{arydshln}
\usepackage[section]{placeins}
\usepackage{natbib}
\usepackage{hyperref}
\usepackage{xr}
\makeatletter
\renewenvironment{abstract}
 {\global\setbox\absbox=\vbox\bgroup
  \hsize=\textwidth
  \noindent\centerline{\large\bfseries Abstract}
  \vspace{1ex}
  \noindent\ignorespaces}
 {\egroup}
\makeatother

\journal{Journal of Political Economy: Microeconomics}

\begin{document}

\begin{frontmatter}

\title{Rebate versus Matching, Again:\\ How Opt-in Reshapes the Effectiveness of Price-Equivalent Subsidies\tnoteref{t1}\\[0.5em]
              \large\normalfont (Last Update: \today)}

\author[aff1]{Shusaku Sasaki\corref{cor1}}
\ead{ssasaki.econ@cider.osaka-u.ac.jp}
 
\author[aff2]{Takunori Ishihara}

\author[aff3]{Hiroki Kato}
 
\cortext[cor1]{Corresponding author.}
 
\address[aff1]{Graduate School of Economics, The University of Osaka, Japan}
\address[aff2]{Faculty of Economics and Business Administration, Kyoto University of Advanced Science, Japan}
\address[aff3]{Center for Infectious Disease Education and Research, The University of Osaka, Japan}

\tnotetext[t1]{We are grateful to Maja Adena, Mark A. Andor, Ren\'{e} Bekkers, Nobuyuki Hanaki, Daniel M. Hungerman, Stephen Knowles, Mark Ottoni-Wilhelm, Sarah Smith, Richard Steinberg, Lukas Tomberg, Joseph Vecci, and Erte Xiao for valuable comments and suggestions. We also thank seminar participants at RWI--Leibniz Institute for Economic Research and the Lilly Family School of Philanthropy at Indiana University, as well as conference participants at the 2023 SPI Conference, the 2025 ESA Asia-Pacific Meeting, the 2025 ESA European Meeting, and the 2025 ARNOVA Annual Conference for helpful discussions. All remaining errors are our own.}

\begin{abstract}
Traditional theory predicts equivalent effects of matching and rebate subsidies at equal prices, yet experiments favor matching. Refinements narrow this gap but retain compulsory assignment; take-up is voluntary in practice. We test this implementation margin in a nationwide, incentivized donation experiment with 2,400 Japanese adults, crossing subsidy type and assignment rule. After equalizing budget constraints and accounting for comprehension, total giving is indistinguishable under compulsory assignment. Under opt-in, the matching advantage re-emerges and nearly quadruples because rebate loses effectiveness. LATE estimates suggest advantageous selection into matching but disadvantageous selection into rebate. Self-selection can reshape the policy ranking of price-equivalent instruments.
\end{abstract}


\end{frontmatter}

\newpage
\section{Introduction}
\label{sec:introduction}

Many public policies employ price subsidies to alter behavior, but their realized impact depends not only on the resulting effective price but also on how the subsidy is implemented. When take-up is voluntary, those who opt in need not be representative of the target population, and the resulting self-selection can amplify or suppress that impact. Much of the existing literature on this margin examines a single program, asking how those who take it up differ from those who do not \citep{alatas2016self,finkelstein2019takeup,ito2023selection}. Comparing two price-equivalent instruments offers a disciplined extension of this approach: holding fixed the mechanical price of the subsidized behavior allows us to focus on how institutional form shapes take-up and realized effectiveness.

Rebate and matching subsidies for charitable giving provide the canonical setting for this question: they are price-equivalent but institutionally distinct schemes whose ranking under compulsory assignment need not carry over to voluntary take-up. This equivalence has made their comparison a natural test of standard theory. A 1:1 matching subsidy is price-equivalent to a 50\% rebate: under a 1:1 matching scheme, an individual who donates 5~USD triggers an additional 5~USD in matching funds, whereas under a 50\% rebate scheme, an individual who donates 10~USD receives a 5~USD refund and therefore faces the same effective out-of-pocket cost. Standard theory thus predicts that the two schemes should yield equivalent outcomes. Controlled experiments have nonetheless repeatedly found stronger responses under matching than under rebate: \citet{eckel2003rebate} first documented higher donation rates and a larger total amount received by the charity under matching, and later studies reported similar matching advantages \citep{eckel2006subsidizing,eckel2008subsidizing,eckel2009encouraging,bekkers2015when}. This evidence established a prominent puzzle under compulsory assignment, because people were assigned to a scheme rather than allowed to choose whether to take it up.

Subsequent experimental refinements have narrowed this puzzle by showing that the advantage of matching over rebate is more modest. Two primary concerns have been carefully addressed: that participants misunderstood the incentive mechanisms \citep{davis2005rebates,davis2005subsidy,davis2006rebate}, and that nominally price-equivalent schemes generate different budget constraints, or feasible sets, under standard third-party designs \citep{lukas2010preference,blumenthal2012subsidizing,sasaki2022experimental}. Most recently, \citet{higgs2025do} equalize these feasible sets within a tax framework and show that the much-cited gap in price responsiveness between rebates and matches disappears, while a modest advantage for matches in the level of giving remains. The meta-analytic evidence also suggests that the matching advantage varies across designs and outcomes, although it is not fully eliminated \citep{clochard2025meta}. In parallel, theory has advanced to explain why price-equivalent schemes generate different responses through the warm-glow and impure-impact channels of the price of giving \citep{hungerman2021impure}. These empirical and theoretical developments may make this comparison appear largely understood. Yet both strands retain compulsory assignment: the take-up decision never enters, and existing theory speaks to the price channel under assignment, not to who would choose the subsidy in the first place.

\begin{figure}[htbp]
  \centering
  \includegraphics[width=\textwidth]{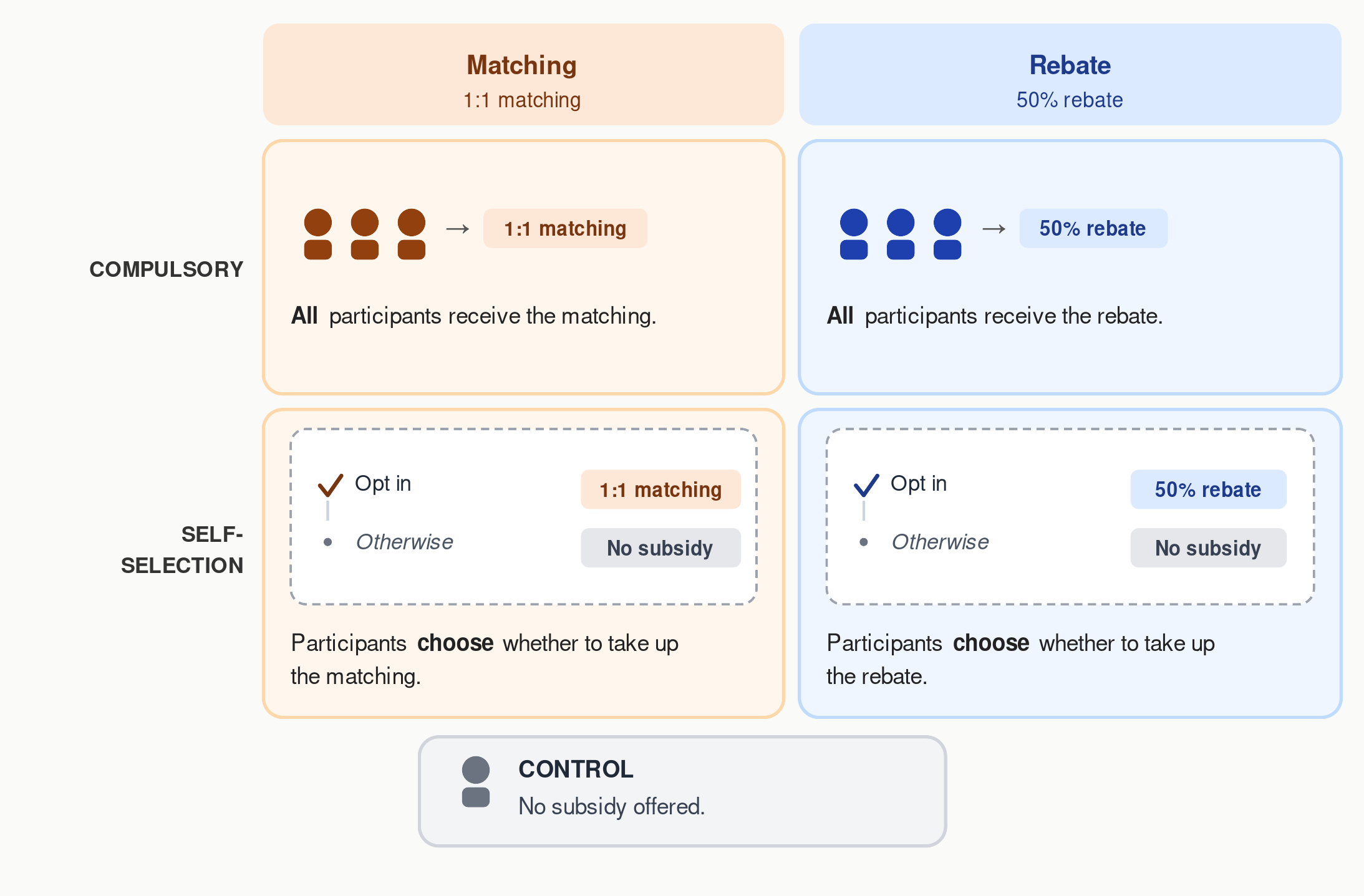}
  \caption{Research Design.}
  \label{fig:design}
\end{figure}

We therefore examine whether this refined compulsory comparison carries over once take-up is voluntary. In that setting, a subsidy's effectiveness depends not only on its average effect under assignment but also on which individuals opt in, what we call the \textit{implementation margin}. We study this margin through a financially incentivized online experiment with a nationwide sample of 2,400 Japanese adults. The experiment employs a $2\times2$ design that crosses the incentive scheme (1:1 matching versus 50\% rebate) with the assignment rule (compulsory versus opt-in), alongside a control group (Figure~\ref{fig:design}). This design allows us to compare each scheme's effect under compulsory assignment with its realized effect when participants decide whether to take it up. Because, in practice, procedures for taking up subsidies can differ across schemes in timing and burden, our controlled experiment standardizes the take-up decision as a symmetric, low-cost choice, so that differences in outcomes under opt-in are not mechanically driven by procedural differences. We also equalize budget constraints and account for comprehension (Section~\ref{sec:design}).

\begin{center}
[\textbf{Figure~\ref{fig:design}} is here.]
\end{center}

This implementation margin is also central to how charitable subsidies operate in practice. In many countries, including the United States, Canada, Australia, and Japan, tax-based subsidies require donors to take steps through tax-filing or related administrative procedures to obtain tax relief; in matching campaigns, donors must identify eligible opportunities and give through designated channels to obtain a match. These requirements make realized policy effects depend on who takes the additional step needed to use the subsidy. This question is at the center of a growing literature on selection into social and environmental programs \citep{alatas2016self,deshpande2019who,domurat2021role,finkelstein2019takeup,ito2023selection,castell2025takeup,ida2026choosing}. In that literature, individuals with larger treatment gains may be more likely to participate, generating advantageous selection, or less likely to participate, generating disadvantageous selection. We bring this selection-on-gains lens to the comparison of price-equivalent charitable subsidies.

Our results reveal a striking re-emergence of the matching advantage once take-up is voluntary. Under the refined compulsory comparison, the matching advantage in the total amount is small and statistically indistinguishable from zero. Under opt-in, however, the gap re-emerges and becomes nearly four times as large as the corresponding compulsory gap. This re-emergence is driven not by matching becoming more effective, but by rebate losing much of its effectiveness once take-up is voluntary. A local average treatment effect (LATE) analysis helps explain this pattern. Under matching, those who opt in appear to be donors for whom the scheme generates relatively large gains in the total amount, consistent with advantageous selection. Under rebate, by contrast, this alignment breaks down: those for whom the scheme would generate the largest gains appear, if anything, less likely to opt in, a pattern suggestive of disadvantageous selection.

Our contribution is neither to re-identify the budget-constraint problem in comparisons of rebate and matching nor to show that matching outperforms rebate under compulsory assignment; both are established in prior work. Rather, we show that voluntary take-up constitutes a distinct implementation margin: even when price equivalence, budget constraints, and comprehension are all accounted for, the donors who select into matching and rebate differ systematically, and this selection can change the schemes' relative effectiveness. The matching side accords with the standard logic of selection on gains. The rebate side does not: those who would give the most appear less likely to opt in. Our exploratory evidence links this pattern to consumption motives, but it lies outside the standard impure-impact framework; we therefore flag it as an open question for future theory rather than a fully identified result. More generally, when price-equivalent instruments are implemented through voluntary take-up, their policy ranking can be governed by who selects in rather than by price alone.

These findings contribute to the literature on the scalability of experimental results, which emphasizes that effects measured under controlled conditions can change once programs scale and participation becomes selective and non-representative \citep{alubaydli2017what,alubaydli2020science,list2022voltage}. We show that this concern can be instrument-specific: introducing the same opt-in decision largely preserves the effectiveness of matching but substantially weakens that of rebate. Methodologically, our compulsory-versus-opt-in design follows the logic of \citet{list2024optimally}'s ``Option C'' experimentation by incorporating a central condition of scale-up into the original trial \citep{fatchen2025using}. The findings also qualify the emerging conclusion that switching between matching and rebate at a common price would do little to increase charitable giving \citep{higgs2025do}: this near-equivalence holds under compulsory assignment but need not extend to voluntary take-up. By isolating this opt-in margin, our results identify a distinct reason why matching may outperform rebate-style subsidies when donors must actively choose to use them.

The remainder of the paper proceeds as follows. Section~\ref{sec:theory} presents the theoretical framework and hypotheses. Section~\ref{sec:design} describes the experimental design. Section~\ref{sec:compulsory} reports the results under compulsory assignment. Section~\ref{sec:selfselection} reports the results under voluntary take-up and examines selection mechanisms. Section~\ref{sec:conclusion} concludes.
\newline \quad

\section{Theoretical Framework and Hypotheses}
\label{sec:theory}

\subsection{Price Channels and Donation Responses}
\label{subsec:theory_price}

Recent theoretical advances have sought to explain why matching and rebate subsidies generate systematically different behavioral responses despite imposing identical prices on donors \citep{hungerman2021impure,higgs2025do}. Using the impure impact model of \citet{hungerman2021impure}, we derive the predictions that guide our empirical analysis. In this framework, donors derive utility from three sources: personal consumption, warm-glow associated with their own checkbook giving amount $g$~\citep{andreoni1989giving,andreoni1990impure}, and the charitable impact of their donation measured by the total amount received by the charity $G$, including any matched funds from a third party. Here, $e<0$ denotes the structural price elasticity of giving, and $\gamma \in [0,1]$ denotes the warm-glow parameter, with $\gamma=0$ indicating no warm-glow preference. The central insight of the model is that, although matching and rebate subsidies can be designed to impose identical out-of-pocket costs on donors, they operate through structurally distinct channels. Formal derivations are presented in \ref{app:theory}.

Specifically, a rebate simultaneously lowers the price of both warm-glow (checkbook giving) and impact (total amount received by the charity), and thus the rebate price elasticity reflects only standard price sensitivity. A matching subsidy, by contrast, lowers the price of impact without affecting the price of warm-glow, so the matching price elasticity additionally reflects warm-glow preferences. A further implication concerns checkbook giving. Under matching, the third-party contribution partially offsets the donor's own giving, a form of strategic substitutability that dampens the checkbook giving response. Under rebate, no such offsetting occurs, and the price reduction translates more directly into increased checkbook giving.

Following \citet{hungerman2021impure}, these structural properties yield two testable predictions. The first concerns checkbook giving:

\begin{quote}
\textbf{Hypothesis 1:} \emph{Regardless of warm-glow preferences, a decrease in the rebate price consistently leads to a larger increase in checkbook giving amounts than an equivalent decrease in the matching price.}
\end{quote}

The second concerns the total amount received by the charity, and its direction depends on the price elasticity of giving $e$ and the warm-glow parameter $\gamma$. Given that empirical estimates typically satisfy $-1 < e < 0$ and $\gamma > 0$ 
(prior-literature averages: $e \approx -0.461$ and $\gamma \approx 0.511$; 
\citealp{eckel2003rebate,eckel2006subsidizing,eckel2008subsidizing,eckel2017comparing,blumenthal2012subsidizing,scharf2015price,hungerman2021impure}), 
matching tends to generate a larger total amount received by the charity 
than an equivalent rebate:

\begin{quote}
\textbf{Hypothesis 2:} \emph{If donors have no warm-glow preference ($\gamma = 0$), the effects of changes in the rebate and matching prices on the total amount received by the charity are the same.}

\emph{If donors have warm-glow preferences ($\gamma > 0$) and are less responsive to donation prices ($-1 < e < 0$), a decrease in the matching price leads to a larger increase in the total amount received by the charity than an equivalent decrease in the rebate price. If they are more responsive to donation prices ($e \leq -1$), a decrease in the rebate price leads to a larger increase in the total amount received by the charity than an equivalent decrease in the matching price.}
\end{quote}
\par\noindent\quad

\subsection{Opt-in Decisions and Selection Effects}
\label{subsec:theory_selection}

While the impure impact model of \citet{hungerman2021impure} was not designed to explicitly incorporate self-selection, its structure nonetheless permits us to derive some predictions about voluntary participation. By examining the model's indirect utility function, we can characterize the incentives governing opt-in decisions and their implications for treatment effects.

The first prediction concerns take-up rates. A donor opts into an incentive scheme when the utility gain from the associated price reduction exceeds the switching cost of participation. When donors have no warm-glow preferences ($\gamma = 0$), the utility gains from rebate and matching are structurally equivalent, implying equal take-up rates. When warm-glow preferences are present ($\gamma > 0$), however, the utility gain under the rebate scheme exceeds that under matching, because a rebate lowers the price of warm-glow as well as impact. This leads to the prediction:

\begin{quote}
\textbf{Hypothesis 3:} \emph{If donors have no warm-glow preferences ($\gamma = 0$), take-up rates are equal across the two schemes. If donors have warm-glow preferences ($\gamma > 0$), the take-up rate under the rebate scheme exceeds that under the matching scheme.}
\end{quote}

The second prediction concerns how self-selection shapes treatment effects. The key mechanism runs through the structure of the utility function itself: since donors derive utility from the total amount received by the charity $G$, the utility gain from taking up either incentive scheme, and thus the motivation to opt in, is directly proportional to how much the total amount received by the charity responds to the price reduction. Under compulsory assignment, the average treatment effect (ATE) captures the mean effect across all participants regardless of individual responsiveness. Under self-selection, however, donors opt in precisely when this utility gain is sufficiently large, meaning those whose total amounts received by the charity respond most strongly to the incentive are disproportionately likely to participate. The treatment effect among actual participants, the Treatment-on-the-Treated (TOT), is therefore expected to exceed the ATE. Correspondingly, the Treatment-on-the-Untreated (TOU), which captures the effect among those who chose not to participate, is expected to fall below the ATE:

\begin{quote}
\textbf{Hypothesis 4:} \emph{For both schemes, individuals who choose to take up the intervention are expected to experience larger increases in total amounts received by the charity than those assigned compulsorily; that is, $\mathrm{TOT} > \mathrm{ATE}$.}
\end{quote}
\par\noindent\quad

\section{Experimental Design}
\label{sec:design}

\subsection{Overview}
\label{sec:overview}

In February 2023, we conducted a financially incentivized online experiment with several design refinements informed by prior literature. The experiment was administered through MyVoice.com Ltd., a company with access to a nationally registered panel of approximately one million adults residing across Japan. Recently, total individual donations in Japan have grown to approximately 2.0 trillion JPY, with growth rates exceeding those in the United States and a giving-to-GDP ratio approaching the levels observed in the United Kingdom and South Korea \citep{GivingJapan2025}. We first conducted a screening survey between February 10 and 15, and then invited screening-survey respondents to the main experiment, which was implemented between February 17 and 21. The main experiment included a quota-balanced sample of 2,400 participants, matched to Japan’s population distribution in terms of age, sex, and residential area (details in \ref{app:sample}).

Participants were financially incentivized through reward points, each valued at 1~JPY. They received fixed fees totaling 65~JPY (15~JPY for the screening survey and 50~JPY for the main experiment) and could earn up to an additional 1,000~JPY based on their decisions during the experimental tasks.\footnote{The fixed fees followed the company's standard compensation schedule. At the February 2023 exchange rate (133~JPY per USD), 65 and 1,000~JPY were worth approximately \$0.5 and \$7.5, respectively.} The screening instrument included questions on demographic attributes, baseline altruism, prior familiarity with the two schemes, and an eight-item comprehension check. The experiment consisted of three main components. First, participants were presented with questions designed to elicit their behavioral characteristics and their prior donation experience. Second, participants were independently and randomly assigned to one of five groups while maintaining the demographic composition of the sample: two matching groups (compulsory and opt-in), two rebate groups (compulsory and opt-in), and a control group, followed by a post-experiment hypothetical donation exercise. Third, participants responded to questions on socioeconomic characteristics, including marital status, number of children, years of education, household income, and place of residence.

The survey items and screenshots of the experimental interface are provided in \ref{app:survey}.
\newline \quad

\subsection{Treatments}
\label{sec:treatments}

The details of each treatment are as follows:

\begin{itemize}
  \item \textbf{1:1 matching (compulsory)}: All participants assigned to this group selected their donation amount under a 1:1 matching scheme, in which the experimenter added an amount equal to each donation. They were informed: \emph{``Our research team will add an amount equal to your donation and deliver it to the charity.''}
  \item \textbf{1:1 matching (opt-in)}: Participants assigned to this group decided whether to opt into the 1:1 matching scheme before making their donation decision. Those who opted in received the same matching terms as the compulsory group; those who chose not to opt in proceeded under the same conditions as the control group.
  \item \textbf{50\% rebate (compulsory)}: All participants assigned to this group selected their donation amount under a 50\% rebate scheme, in which the experimenter refunded 50\% of the donated amount to the participant. They were informed: \emph{``Our research team will cover 50\% of your donation amount and refund it to you.''}
  \item \textbf{50\% rebate (opt-in)}: Participants assigned to this group decided whether to opt into the 50\% rebate scheme before making their donation decision. Those who opted in received the same rebate terms as the compulsory group; those who chose not to opt in proceeded under the same conditions as the control group.
\end{itemize}

In the opt-in conditions, the default setting was set to ``not use'' the incentive, introducing a minor and symmetric switching cost for participants who chose to opt in. By applying the identical default setting to both the matching and rebate opt-in conditions, we enable a clean comparison of how the opportunity to opt in influences the effectiveness of each incentive.
\newline \quad

\subsection{Donation Decision and Outcomes}
\label{sec:donation}

Participants were informed that one in ten would be randomly selected to receive an additional reward worth 1,000~JPY. A recent meta-analysis shows that random-payment designs elicit behavior comparable to all-payment in dictator-game contexts \citep{umer2023effectiveness}. They were then asked to decide how much of this amount they would donate to a social contribution project,\footnote{The charity was named the Nature Conservation Society of Japan, a government-certified public interest foundation dedicated to protecting endangered natural environments in Japan.} conditional on receiving the 1,000~JPY. For selected participants, the donation decision was binding and carried out as stated.

Our primary outcomes are ``\emph{checkbook giving},'' defined as the amount the participant chooses to donate from the endowment in 100~JPY increments, and ``\emph{the total amount received by the charity},'' which includes any matched funds under the matching scheme. We use 100~JPY increments to discretize the choice set, motivated by the unit donation framing of \citet{diederich2022subsidizing}. The donor's cost differs structurally between schemes (100~JPY under matching versus 50~JPY under rebate) at both the extensive and intensive margins. This asymmetry is inherent to rebate-matching comparisons whenever donations have a minimum positive unit, but the unit donation framing makes it salient. This asymmetry could favor rebate, which makes the test of matching superiority more conservative.

The experimental screen automatically calculated and displayed ``the checkbook amount,'' ``the reward to self,'' and ``the total amount received by the charity'' after participants chose their initial amount. Participants were required to confirm their selection by clicking a button before it was recorded, a step designed to minimize confusion about the incentive mechanisms and mitigate concerns about the isolation effect documented in prior studies \citep{davis2005rebates,davis2005subsidy,davis2006rebate}.
\newline \quad

\subsection{Budget Constraint Differences between Rebate and Matching}
\label{sec:budget}

As illustrated in Appendix Figure~\ref{fig:budget_constraint}, there exists a structural difference in the budget constraint line between the matching and rebate schemes \citep{lukas2010preference,blumenthal2012subsidizing,sasaki2022experimental}. Suppose a donor is given an endowment of $1,000$~JPY and decides how much to donate. Under the control condition, if the donor initially selects $g$ JPY, the total amount received by the charity is $g$ JPY and the reward to self is $(1,000 - g)$ JPY, so the maximum total amount received by the charity is $1,000$ JPY. Under the 50\% rebate scheme, the total amount received by the charity is $g$ JPY but the reward to self is $(1,000 - 0.5g)$ JPY, so the maximum out-of-pocket expenditure is $500$ JPY and the maximum total amount received by the charity is $1,000$ JPY. Under the 1:1 matching scheme, the total amount received by the charity is $2g$ JPY and the reward to self remains $(1,000 - g)$ JPY, so the maximum total amount received by the charity is $2,000$ JPY.

That is, although matching and rebate impose identical prices on donors, they differ in the maximum total amount received by the charity. Without accounting for this difference, the treatment effect of matching on the total amount received by the charity would be overestimated.

Following \citet{blumenthal2012subsidizing}, we designed the experiment to account for this difference. Participants faced two donation questions, each with the same budget of $1,000$~JPY but different upper limits on the initial donation: $1,000$~JPY and $500$~JPY, presented in randomized order (see Appendix Table~\ref{tab:budget_example}). In the opt-in conditions, participants decided whether to opt in before completing each question. By using responses from the $500$~JPY upper-limit question for the matching group and the $1,000$~JPY upper-limit question for the rebate and control groups, we equalize the maximum total amount received by the charity across conditions. Appendix Table~\ref{tab:budget_example} illustrates this design using an example where each group selects 80\% of the upper limit as their initial donation.
\newline \quad

\subsection{Ethics and Pre-registration}
\label{subsec:ethics_prereg}

This study was approved by the ethics committee of The University of Osaka, Japan (2022CRER0120-2). The experimental design and procedures were also pre-registered with the AEA RCT Registry \citep{sasaki2023rebate}. We follow the pre-analysis plan; its details and any extensions are provided in \ref{app:preregi}.
\newline \quad

\subsection{Balance Check}
\label{sec:balance}

Table~\ref{tab:balance} reports means and standard deviations of key baseline characteristics for each of the five groups, including sex, age, family structure, years of education, household income, place of residence, and baseline altruism.\footnote{Baseline altruism was elicited in the screening survey as the amount, in 100 JPY increments, respondents would hypothetically donate from a 1,000 JPY windfall to an environmental conservation activity.} No statistically significant differences are found across groups, confirming that randomization achieved balance on observable characteristics.

\begin{center}
[\textbf{Table~\ref{tab:balance}} is here]
\end{center}
\par\noindent\quad

\section{Compulsory Treatment Effects}
\label{sec:compulsory}

\subsection{Average Treatment Effects}
\label{sec:ate}

\begin{center}
[\textbf{Table~\ref{tab:compulsory}} is here]
\end{center}

Building on refinements proposed in the literature following \citet{eckel2003rebate}, we first examine whether the apparent superiority of matching over rebate survives two adjustments that subsequent studies have identified as important: accounting for differences in budget constraints \citep{lukas2010preference,blumenthal2012subsidizing,sasaki2022experimental} and controlling for participants' misunderstandings of the mechanisms \citep{davis2005rebates,davis2005subsidy,davis2006rebate}.

Without adjusting for budget constraints (Columns~1 and~2 of Table~\ref{tab:compulsory}), the results mirror the prior literature: under compulsory assignment, the 1:1 matching outperforms the 50\% rebate in the total amount received by the charity by 271~JPY. Consistent with Hypothesis~1, the rebate raises checkbook giving more than the matching. Once budget constraints are equalized (Columns~3 and~4), the matching advantage in the total amount narrows sharply, from 271~JPY to 59.4~JPY.

The picture becomes even clearer when we further restrict the sample to participants with a high level of understanding of the mechanisms (six or more correct answers; Columns~5 and~6). Among this group, the compulsory 1:1 matching raises the total amount received by the charity by 236~JPY and the 50\% rebate by 210~JPY. The resulting gap of only 26~JPY, a tenfold reduction from the 271~JPY observed without any adjustments, is statistically indistinguishable from zero ($p = .351$). The remaining participants, those with lower understanding (Columns~7 and~8), continue to exhibit a significant matching advantage (205.8~JPY versus 104.4~JPY; $p < .01$), suggesting that, at least in our sample, misunderstanding or confusion about the mechanisms contributes to the apparent superiority of matching.

The central finding from the compulsory conditions is that, once differences in budget constraints and participants' misunderstandings are both accounted for, the previously observed advantage of matching over rebate in the total amount received by the charity is no longer detectable. This result is consistent with \citet{davis2006rebate}, who showed that the matching advantage disappears when the total amount received by the charity is made most salient to participants. To summarize in terms of our hypotheses: Hypothesis~1 is supported, as the rebate consistently produces larger checkbook giving than the matching across specifications. For the total amount, the absence of a detectable matching advantage in the high-comprehension sample is consistent with the $\gamma \approx 0$ case of Hypothesis~2, pointing to a warm-glow parameter close to zero.
\newline \quad

\subsection{Price Elasticity and Warm-Glow Estimation}
\label{sec:elasticity}

We directly estimate the price elasticity of giving and the warm-glow parameter to verify whether the latter is indeed close to zero. Using the Poisson regression approach of \citet{chen2024logs}, we model checkbook giving as:
\begin{equation}
g = \exp\!\left( \beta_{0} + \beta_{1} \cdot \ln(p_{M}) + \beta_{2} \cdot \ln(p_{R}) \right),
\label{eq:poisson}
\end{equation}
where $\beta_{1}$ and $\beta_{2}$ can be directly interpreted as the price elasticities with respect to the matching and rebate prices, respectively. We obtain a matching price elasticity of $0.306~(\mathrm{S.E.} = 0.098)$ and a rebate price elasticity of $-0.632~(\mathrm{S.E.} = 0.084)$. Mapping these to the structural parameters of Section~\ref{sec:theory}, where the rebate price elasticity equals $e$ and the matching price elasticity equals $(1+e)(1-\gamma)$, yields $e = -0.632~(\mathrm{S.E.} = 0.084)$ and $\gamma = 0.168~(\mathrm{S.E.} = 0.170)$.

The estimated price elasticity satisfies $-1 < e < 0$, 
close to the prior-literature average of $-0.461$ 
\citep{eckel2003rebate,eckel2006subsidizing,eckel2008subsidizing,eckel2017comparing,blumenthal2012subsidizing,scharf2015price,hungerman2021impure}. 
By contrast, the warm-glow parameter is smaller than the 
prior-literature average of $0.511$ from the same set of studies, 
and importantly, the 95\% confidence interval of $[-0.164, 0.501]$ 
is consistent with $\gamma = 0$ at conventional levels.

These estimates confirm and reinforce the message from Section~\ref{sec:ate}. The warm-glow parameter close to zero implies that the utility gains from the two schemes are structurally equivalent, consistent with the finding that, when budget constraints are equalized and the sample is restricted to participants who correctly understand the mechanisms, the matching advantage over rebate in the total amount received by the charity is no longer detectable. They appear to respond on average in nearly the same way to matching and rebate.
\newline \quad

\section{Treatment Effects with Self-selection}
\label{sec:selfselection}

The convergence documented in Section~\ref{sec:compulsory} extends to take-up rates. In the high-comprehension sample, 65.3\% chose to take up the matching treatment and 63.7\% the rebate ($p = 0.699$), a near-equivalence consistent with the $\gamma \approx 0$ prediction of Hypothesis~3.

What happens next, however, is striking. Despite this convergence at every turn, in compulsory treatment effects, in estimated parameters, and in take-up rates, the introduction of self-selection dramatically restores the superiority of matching over rebate. The gap in the total amount received by the charity, which had been statistically indistinguishable from zero under compulsory assignment, expands substantially when individuals are free to choose whether to participate. This re-emergence is driven not by matching becoming more effective, but by the rebate losing much of its effectiveness under self-selection. This suggests that the option to opt in works very differently under each scheme.
\newline \quad

\subsection{Treatment Effects: ITT and LATEs}
\label{sec:itt-late}

Table~\ref{tab:selfsel} reports the intent-to-treat (ITT) estimates for the self-selection conditions using the high-comprehension sample. For checkbook giving, the advantage of the rebate weakens under self-selection: the 1:1 matching slightly decreased checkbook giving by 41.1~JPY relative to the control group, while the 50\% rebate increased it by 113~JPY, roughly half the ATE observed under compulsory assignment. The gap between the two treatments narrows accordingly.

For the total amount received by the charity, the picture is reversed. The 1:1 matching increased the total amount by 210.7~JPY, close to its compulsory ATE, while the 50\% rebate increased the total amount by only 113~JPY, barely half of its compulsory counterpart. The difference between the two treatments expands to 97.7~JPY, nearly four times the corresponding gap under compulsory assignment.

\begin{center}
[\textbf{Table~\ref{tab:selfsel}} is here]
\end{center}

To understand why, we follow the LATE framework of \citet{fowlie2021default} and decompose treatment effects by participants' take-up status under voluntary assignment. The Treatment-on-the-Treated (TOT) is the effect for participants who would take up the subsidy when offered under opt-in (Takers), whereas the Treatment-on-the-Untreated (TOU) is the effect for those who would not (Non-takers). The TOT is identified by comparing the control group with the opt-in group, using random assignment to the opt-in offer as an instrument for actual subsidy take-up. The TOU is recovered by dividing the difference between the compulsory ATE and the opt-in ITT by the share of non-takers.

\begin{figure}[htbp]
  \centering
  \includegraphics[
    width=\textwidth,
    trim=0.80in 0.70in 0.70in 0.55in, clip
  ]{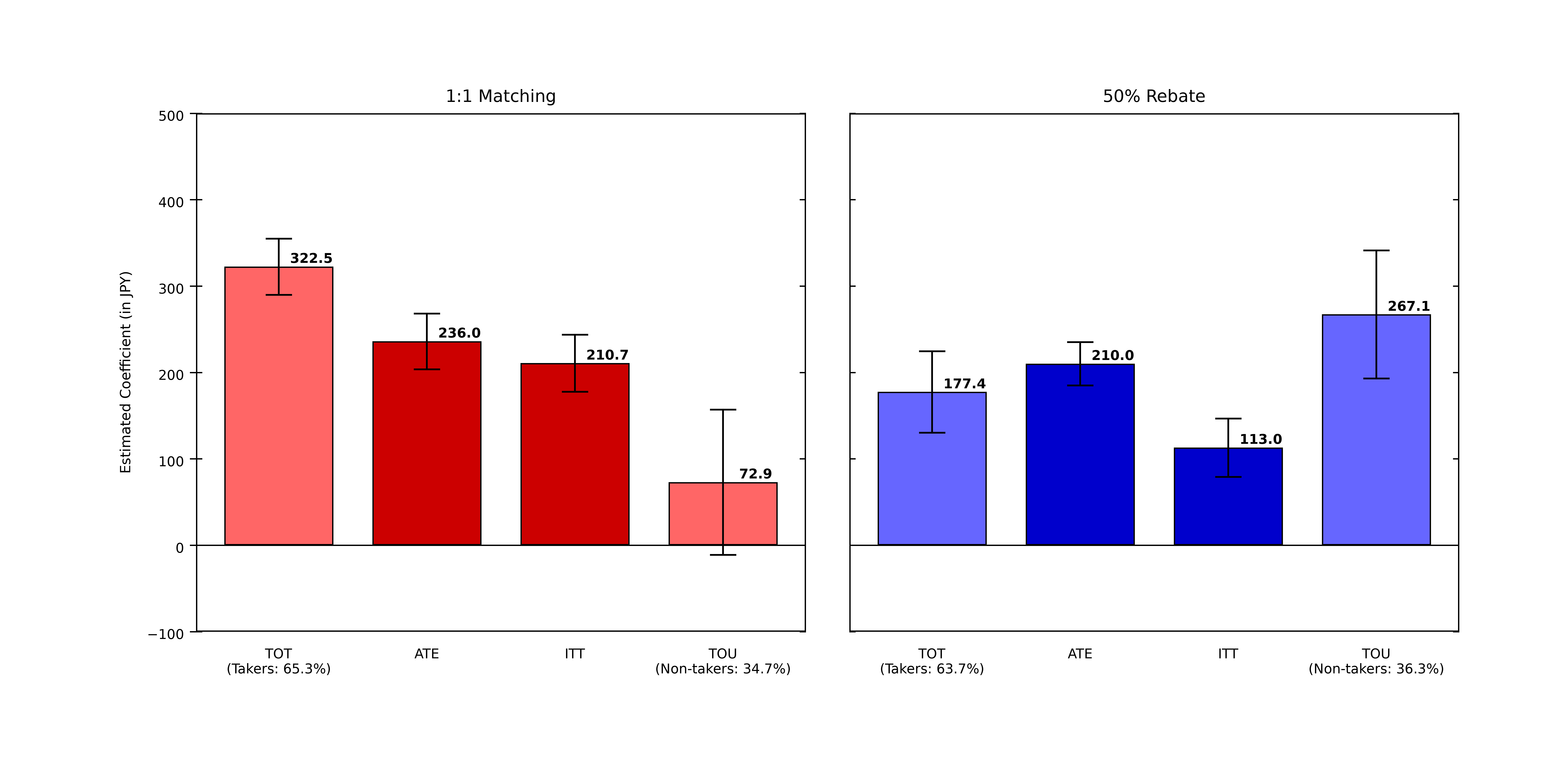}
  \caption{Treatment Effects on Total Amount Received by Charity.}
  \begin{minipage}{0.95\textwidth}
    \footnotesize
    \emph{Notes:} Treatment effects are reported in JPY. At the time of the experiment, 1 USD was approximately 133 JPY. ATE denotes the effect of compulsory assignment to the corresponding subsidy scheme relative to the control group. ITT denotes the effect of offering the subsidy under voluntary take-up relative to the control group. TOT denotes the treatment effect for participants who would take up the subsidy under voluntary take-up; TOU denotes the treatment effect for those who would not. Vertical bars indicate cluster-robust standard errors at the regional level. The analysis uses the high-comprehension sample, defined as participants scoring above the median on the eight-item matching-and-rebate comprehension check (six or more correct answers).
  \end{minipage}
  \label{fig:late1}
\end{figure}

\begin{center}
[\textbf{Figure~\ref{fig:late1}} is here]
\end{center}

The resulting TOT and TOU estimates show that selection patterns differ sharply across the two schemes. Under the 1:1 matching, the TOT is 322.5~JPY, substantially larger than the ATE of 236.0~JPY ($p = 0.003$), indicating that Takers were precisely those who increased the total amount received by the charity most. The TOU is only 72.9~JPY and statistically insignificant, suggesting that those unlikely to increase their giving chose not to participate. Selection under matching is thus advantageous: those with the most to gain are the most likely to participate.

Under the 50\% rebate, however, the pattern points in the opposite direction.  The TOT of 177.4~JPY does not differ significantly from the ATE of 210.0~JPY ($p = 0.432$). The point estimates follow the order TOT $<$ ATE $<$ TOU ($267.1$~JPY). This ordering would arise under disadvantageous selection, in which individuals who would have increased the total amount received by the charity most are less likely to take up the rebate.

Taken together, these asymmetric selection patterns explain the reversal documented in Table~\ref{tab:selfsel}. The act of choosing treats the two schemes very differently: under matching, advantageous selection amplifies the ITT effect, while under rebate, disadvantageous selection suppresses it. These patterns are consistent with Hypothesis~4 for matching. The TOT significantly exceeds the ATE, consistent with the prediction that those who stand to gain most are the most likely to participate. However, the opposite holds for the rebate: those who would have benefited most did not opt in, a pattern at odds with the prediction.
\newline \quad

\subsection{Robustness Checks}
\label{sec:robustness}

Our findings are robust to additional checks (Appendix 
Tables~\ref{tab:robustness_selfselection} and~\ref{tab:mht_corrections}). First, covariate-adjusted OLS estimates and a two-limit Tobit model yield matching advantages under self-selection of 94~JPY and 100~JPY, respectively ($p = 0.006$ and $p < 0.001$), while the corresponding gaps under compulsory assignment are much smaller and not significant at the 5\% level. Second, the extensive-margin estimates are directionally consistent with the checkbook-giving results: the rebate advantage in the probability of a positive donation declines from 8.3 percentage points under compulsory assignment ($p = 0.009$) to 0.4 percentage points under self-selection ($p = 0.861$). Third, a post-experiment hypothetical exercise with a tenfold-larger endowment (10{,}000~JPY) also preserves the larger matching advantage under self-selection (849~JPY, $p = 0.005$), whereas the compulsory gap remains insignificant ($p = 0.443$). Finally, although the pre-analysis plan did not pre-specify a multiple-testing correction, the matching-versus-rebate difference in the total amount under self-selection remains significant under both the Romano--Wolf correction \citep{RomanoWolf2005} ($p = 0.004$) and the List--Shaikh--Xu correction \citep{ListShaikhXu2019} ($p = 0.010$), applied to the family of four contrasts on the two primary outcomes across regimes.
\newline \quad

\subsection{Exploring Selection Mechanisms}
\label{sec:mechanisms}

We conduct two exploratory analyses designed to shed light on the mechanisms behind the asymmetric selection patterns documented in Section~\ref{sec:itt-late}. First, we regress take-up on a set of baseline characteristics, including altruism scores and other covariates, separately for each scheme. The results in Table~\ref{tab:takeup_regression} reveal a clear asymmetry: individuals with higher altruism are significantly more likely to take up the matching treatment, whereas no significant effect is found for the rebate. The matching pattern is consistent with Hypothesis~4: utility gains from participation increase with the altruism parameter, making more altruistic individuals more likely to opt in, which explains why the TOT for donations substantially exceeds the ITT under matching. However, the same logic does not hold for the rebate: altruism does not predict take-up, suggesting that opt-in decisions under rebate are governed by factors outside the model.

\begin{center}
[\textbf{Table~\ref{tab:takeup_regression}} is here]
\end{center}

\begin{figure}[htbp]
  \centering
  \includegraphics[
    width=\textwidth,
    trim=0.80in 0.70in 0.70in 0.55in, clip
  ]{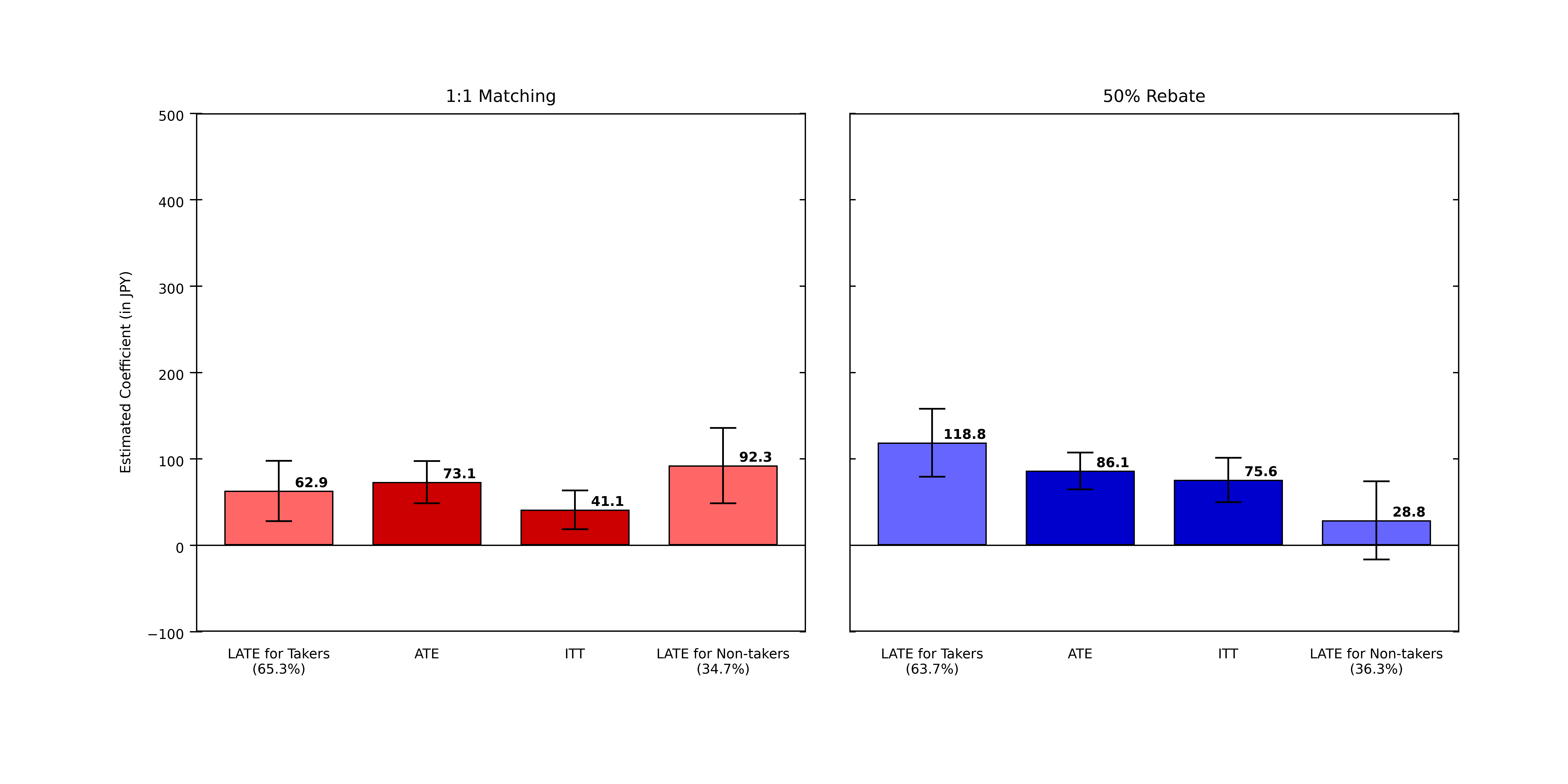}
  \caption{Treatment Effects on Consumption (Reward to Self).}
  \begin{minipage}{0.95\textwidth}
    \footnotesize
    \emph{Notes:} Treatment effects are reported in JPY. At the time of the experiment, 1 USD was approximately 133 JPY. ATE denotes the effect of compulsory assignment to the corresponding subsidy scheme relative to the control group. ITT denotes the effect of offering the subsidy under voluntary take-up relative to the control group. TOT denotes the treatment effect for participants who would take up the subsidy under voluntary take-up; TOU denotes the treatment effect for those who would not. Vertical bars indicate cluster-robust standard errors at the regional level. The analysis uses the high-comprehension sample, defined as participants scoring above the median on the eight-item matching-and-rebate comprehension check (six or more correct answers).
  \end{minipage}
  \label{fig:late2}
\end{figure}

Second, among the non-altruistic motives that could drive take-up under rebate, one that our data allow us to test is the desire to capture the refund for personal consumption. We estimate ATEs, ITTs, and LATEs using consumption (reward to self) as the outcome variable. Figure~\ref{fig:late2} shows that, under matching, the TOT for consumption falls below the ATE, consistent with Takers being altruistically motivated: they donate more and retain less for themselves. Under rebate, the pattern is reversed: the TOT substantially exceeds both the ATE and the TOU, indicating that Takers disproportionately increase their own consumption. Together, these findings suggest that donors opt into the rebate primarily to capture the refund rather than to increase charitable giving, providing evidence consistent with the interpretation that consumption-oriented considerations contribute to take-up under rebate

\begin{center}
[\textbf{Figure~\ref{fig:late2}} is here]
\end{center}
\par\noindent\quad

\section{Discussion and Conclusions}
\label{sec:conclusion}

Rebate and matching subsidies are price-equivalent instruments whose relative effectiveness has long posed a puzzle in charitable giving. Prior experimental work has focused on compulsory assignment, leaving open whether the comparison carries over when donors must choose whether to use the subsidy. Using a nationwide sample of Japanese adults in a financially incentivized online experiment, we implement a $2\times2$ design crossing incentive scheme, 1:1 matching versus 50\% rebate, with assignment rule, compulsory assignment versus opt-in, alongside a control group. We equalize switching costs and budget constraints and account for participants’ understanding of the schemes. Once these refinements are incorporated, the matching advantage in the total amount is no longer detectable under compulsory assignment; however, this near-equivalence does not carry over to voluntary take-up. When participants can choose whether to use the subsidy, the matching advantage re-emerges and becomes nearly four times as large as the corresponding compulsory gap. This re-emergence reflects not improved performance of matching but a substantial loss of effectiveness under rebate.

This asymmetry stems from how donors self-select into each scheme. Under matching, donors who stand to gain more from the subsidy are more likely to opt in, generating advantageous selection in which the Treatment-on-the-Treated exceeds the Average Treatment Effect. Under rebate, by contrast, this alignment breaks down: the most responsive donors appear less likely to participate, producing a pattern suggestive of disadvantageous selection. Altruism significantly predicts take-up under matching but not under rebate. Under our estimated warm-glow parameter of $\gamma \approx 0$, the impure-impact framework predicts advantageous selection of similar strength in both schemes; the matching pattern is consistent with this, but the rebate pattern is not. The opt-in decision under rebate therefore appears to be governed by considerations outside this standard framework.

One potential concern is whether the rebate anomaly is an artifact of our $\gamma \approx 0$ framework. This does not seem to be the case. Our point estimate and the literature average both place $\gamma$ above zero, motivating $\gamma > 0$ as a natural robustness check. However, under $\gamma > 0$, the framework predicts higher take-up under rebate than under matching and advantageous selection in both schemes. Neither prediction is borne out: take-up is not higher under rebate, and rebate exhibits disadvantageous rather than advantageous selection. Our central message that opt-in decisions under rebate are governed by considerations outside the standard model is therefore robust to the value of the warm-glow parameter.

What lies outside the impure-impact model? This standard framework predicts advantageous selection in both schemes, and under our estimated warm-glow parameter of $\gamma \approx 0$, selection should be similar in strength across matching and rebate. Yet this prediction fails for the rebate, pointing to psychological costs or non-altruistic considerations that existing models do not capture. Our findings thus reopen a question long regarded as settled and sharpen the target for future theory. Two pieces of evidence suggest that consumption-oriented motives may shape participation decisions under rebate in ways that current models cannot accommodate. As shown in Section~\ref{sec:mechanisms}, some donors under the rebate scheme appear to place greater weight on personal consumption than on charitable impact when deciding whether to participate. \citet{chan2022perception} likewise show that donors perceive rebates as involving a personal reward and therefore as less generous than matching subsidies. Closing these theoretical gaps is essential for developing more accurate behavioral models and for designing more effective charitable giving policies.

The policy implication of our study is that, when subsidies are implemented on a voluntary basis, matching tends to be the more effective instrument. This bears on ongoing policy debates over charitable tax relief. The United Kingdom's Gift Aid, for example, converts tax relief into a charity-side top-up rather than a donor-side rebate, reflecting a broader concern that conventional rebate-style deductions may do too little to generate additional giving.

One limitation concerns external validity. Our experiment standardizes opt-in as a symmetric, low-cost click, whereas real-world rebates impose substantially higher procedural burdens: donors must file post-donation claims, potentially after the donation amount has already been decided, and face considerable administrative costs. Consistent with this, \citet{eckel2017comparing} document a larger acceptance gap (rebate 38\% vs.\ matching 73\%) in a naturalistic mail solicitation, suggesting that frictions like differential awareness may amplify the matching advantage beyond the selection on motivation our study isolates. However, other recent evidence suggests that matching subsidies could lose effectiveness when donors can avoid the solicitation, since the social pressure through which matching operates can then be sidestepped \citep{gangadharan2026notjust}. Therefore, how these opposing forces balance out in real-world contexts should be examined in future research, and will likely remain a central question in the study of charitable subsidies.

\newpage
\section*{Funding}
This research is financially supported by the Japan Society for the Promotion of Science {[}Grant Numbers: 19K13722 and 24K00264 (S.~Sasaki){]} and Japan Science and Technology Agency {[}Grant Number: JPMJPR21R4 (S.~Sasaki){]}.

\section*{Declaration of competing interests}
The authors declare no conflicts of interest.

\newpage
\bibliographystyle{elsarticle-harv}
\bibliography{references}


\newpage

\newpage

\begin{table}[htbp]
  \centering
  \caption{Balance Check}
  \label{tab:balance}
  \begin{threeparttable}
  \small
  \setlength{\tabcolsep}{4pt}
  \begin{tabular}{lcccccc}
    \toprule
     & Control & \multicolumn{2}{c}{1:1 matching} & \multicolumn{2}{c}{50\% rebate} &  \\
    \cmidrule(lr){3-4}\cmidrule(lr){5-6}
     & (1) & (2) Compulsory & (3) Opt-in & (4) Compulsory & (5) Opt-in & $p$-value \\
    \midrule
    Female                & 0.502    & 0.502    & 0.502    & 0.502    & 0.502    & 1.000 \\
                          & [0.501]  & [0.501]  & [0.501]  & [0.501]  & [0.501]  &       \\
    \addlinespace[4pt]
    Age                   & 46.398   & 46.369   & 45.946   & 46.108   & 46.144   & 0.983 \\
                          & [13.042] & [13.168] & [13.285] & [13.201] & [13.138] &       \\
    \addlinespace[4pt]
    Married               & 0.550    & 0.492    & 0.517    & 0.546    & 0.517    & 0.350 \\
                          & [0.498]  & [0.500]  & [0.500]  & [0.498]  & [0.500]  &       \\
    \addlinespace[4pt]
    Having children       & 0.248    & 0.237    & 0.258    & 0.283    & 0.246    & 0.530 \\
                          & [0.432]  & [0.426]  & [0.438]  & [0.451]  & [0.431]  &       \\
    \addlinespace[4pt]
    Educational years     & 14.493   & 14.748   & 14.634   & 14.692   & 14.760   & 0.265 \\
                          & [2.126]  & [2.093]  & [2.115]  & [2.049]  & [2.034]  &       \\
    \addlinespace[4pt]
    Household income      & 587.961  & 566.012  & 581.190  & 619.613  & 609.509  & 0.126 \\
    \quad (10,000 JPY)    & [368.899]& [332.549]& [330.660]& [373.774]& [359.703]&       \\
    \addlinespace[4pt]
    No income information & 0.198    & 0.165    & 0.169    & 0.171    & 0.185    & 0.641 \\
                          & [0.399]  & [0.371]  & [0.375]  & [0.377]  & [0.389]  &       \\
    \addlinespace[4pt]
    Living in urban areas & 0.554    & 0.556    & 0.563    & 0.560    & 0.571    & 0.988 \\
                          & [0.498]  & [0.497]  & [0.497]  & [0.497]  & [0.495]  &       \\
    \addlinespace[4pt]
    Baseline altruism     & 304.167  & 280.417  & 308.333  & 318.958  & 338.750  & 0.120 \\
    \quad (JPY)           & [349.206]& [334.626]& [341.575]& [344.832]& [354.627]&       \\
    \midrule
    Number of observations & 480     & 480      & 480      & 480      & 480      &       \\
    \bottomrule
  \end{tabular}
  \begin{tablenotes}
    \footnotesize
    \item \emph{Notes:} Standard deviations are reported in brackets. Female, Married, Having children, No income information, and Living in urban areas are binary indicators. For respondents who did not report household income, we imputed the sample mean and included a separate indicator (``No income information''). Baseline altruism was measured in the screening survey as the amount that respondents would hypothetically donate from a 1{,}000~JPY windfall to an environmental conservation activity. At the time of the experiment, 1 USD was approximately 133 JPY. The $p$-value column reports the $F$-test of joint equality of means across the five experimental groups, obtained by regressing each variable on group indicators. None of these tests rejects equality at the 5\% level.
  \end{tablenotes}
  \end{threeparttable}
\end{table}

\newpage
\begin{table}[htbp]
  \centering
  \caption{Compulsory Treatment Effects}
  \label{tab:compulsory}
  \begin{threeparttable}
  \scriptsize
  \setlength{\tabcolsep}{3pt}
  \begin{tabular}{l*{8}{c}}
    \toprule
     & (1) & (2) & (3) & (4) & (5) & (6) & (7) & (8) \\
    \cmidrule(lr){2-3}\cmidrule(lr){4-9}
    Exp.\ design: & \multicolumn{2}{c}{Without budget adj.} & \multicolumn{6}{c}{With budget adjustment} \\
    \cmidrule(lr){2-3}\cmidrule(lr){4-5}\cmidrule(lr){6-7}\cmidrule(lr){8-9}
    Sample: & \multicolumn{2}{c}{Full sample} & \multicolumn{2}{c}{Full sample} & \multicolumn{2}{c}{High-comprehension} & \multicolumn{2}{c}{Low-comprehension} \\
    \cmidrule(lr){2-3}\cmidrule(lr){4-5}\cmidrule(lr){6-7}\cmidrule(lr){8-9}
    Dep.\ var.: & Checkbook & Total & Checkbook & Total & Checkbook & Total & Checkbook & Total \\
    \midrule
    1:1 matching        & 43.125     & 435.208*** & $-62.708$*** & 223.542*** & $-73.089$** & 235.964*** & $-57.053$* & 205.817*** \\
    Compulsory (ATE)    & (27.576)   & (45.657)   & (17.736)     & (25.365)   & (24.462)    & (32.237)   & (27.197)   & (33.046)   \\
    \addlinespace[4pt]
    50\% rebate         & 164.167*** & 164.167*** & 164.167***   & 164.167*** & 209.983***  & 209.983*** & 104.414**  & 104.414**  \\
    Compulsory (ATE)    & (17.777)   & (17.777)   & (17.777)     & (17.777)   & (25.075)    & (25.075)   & (33.524)   & (33.524)   \\
    \addlinespace[8pt]
    Constant            & 348.958*** & 348.958*** & 348.958***   & 348.958*** & 382.143***  & 382.143*** & 319.922*** & 319.922*** \\
                        & (12.700)   & (12.700)   & (12.700)     & (12.700)   & (20.595)    & (20.595)   & (24.188)   & (24.188)   \\
    \midrule
    Match (ATE) $-$ Rebate (ATE) & $-121.042$ & 271.041 & $-226.875$ & 59.375 & $-283.072$ & 25.981 & $-161.467$ & 101.403 \\
    \addlinespace[2pt]
    \quad $p$-value     & $<$0.001   & $<$0.001   & $<$0.001     & 0.001      & $<$0.001    & 0.351      & $<$0.001   & 0.003      \\
    \addlinespace[2pt]
    Observations        & 1{,}440    & 1{,}440    & 1{,}440      & 1{,}440    & 721         & 721        & 719        & 719        \\
    \addlinespace[2pt]
    $R^{2}$             & 0.039      & 0.123      & 0.089        & 0.062      & 0.143       & 0.076      & 0.044      & 0.051      \\
    \bottomrule
  \end{tabular}
  \begin{tablenotes}
    \footnotesize
    \item \emph{Notes:} Cluster-robust standard errors at the regional level are reported in parentheses. *** $p<0.01$, ** $p<0.05$, * $p<0.1$. ``Checkbook'' refers to the participant's own donation amount; ``Total'' refers to the total amount received by the charity, including matching funds where applicable. Treatment effects are reported in JPY. At the time of the experiment, 1 USD was approximately 133 JPY. ``With budget adjustment'' uses upper-limit responses that equalize the maximum total amount received by the charity across schemes (see Section~\ref{sec:budget}). The high-comprehension sample comprises participants with six or more correct answers on the eight-item comprehension check; the low-comprehension sample comprises those with five or fewer. The matching advantage in the total amount received by the charity attenuates as the budget constraint and comprehension are progressively accounted for, becoming statistically indistinguishable from the rebate effect in Column (6).
  \end{tablenotes}
  \end{threeparttable}
\end{table}

\newpage
\begin{table}[htbp]
  \centering
  \caption{Treatment Effects under Self-selection}
  \label{tab:selfsel}
  \begin{threeparttable}
  \small
  \begin{tabular}{l*{4}{c}}
    \toprule
     & (1) & (2) & (3) & (4) \\
    \cmidrule(lr){2-3}\cmidrule(lr){4-5}
    Sample: & \multicolumn{2}{c}{High-comprehension} & \multicolumn{2}{c}{Low-comprehension} \\
    \cmidrule(lr){2-3}\cmidrule(lr){4-5}
    Dependent variable: & Checkbook & Total & Checkbook & Total \\
    \midrule
    1:1 matching         & $-41.107$*  & 210.686***  & $-64.027$    & 44.707      \\
    Opt-in (ITT)         & (22.404)    & (32.956)    & (37.759)     & (51.576)    \\
    \addlinespace[4pt]
    50\% rebate          & 112.959***  & 112.959***  & 47.312**     & 47.312**    \\
    Opt-in (ITT)         & (33.737)    & (33.737)    & (19.691)     & (19.691)    \\
    \midrule
    1:1 matching         & $-73.089$** & 235.964***  & $-57.053$*   & 205.817***  \\
    Compulsory (ATE)     & (24.462)    & (32.237)    & (27.197)     & (33.046)    \\
    \addlinespace[4pt]
    50\% rebate          & 209.983***  & 209.983***  & 104.414**    & 104.414**   \\
    Compulsory (ATE)     & (25.075)    & (25.075)    & (33.524)     & (33.524)    \\
    \midrule
    Constant             & 382.143***  & 382.143***  & 319.922***   & 319.922***  \\
                         & (20.595)    & (20.595)    & (24.188)     & (24.188)    \\
    \midrule
    Match (ITT) $-$ Rebate (ITT) & $-154.066$ & 97.727  & $-111.339$ & $-2.605$   \\
    \quad $p$-value      & $<$0.001      & 0.007      & 0.003       & 0.951      \\
    \addlinespace[4pt]
    Match (ATE) $-$ Rebate (ATE) & $-283.072$ & 25.981 & $-161.467$  & 101.403    \\
    \quad $p$-value      & $<$0.001    & 0.351      & $<$0.001     & 0.003      \\
    \addlinespace[4pt]
    Take-up rate of 1:1 matching & 65.3\% & 65.3\%   & 30.6\%       & 30.6\%      \\
    Take-up rate of 50\% rebate  & 63.7\% & 63.7\%   & 36.6\%       & 36.6\%      \\
    \addlinespace[4pt]
    Observations         & 1{,}217     & 1{,}217     & 1{,}183      & 1{,}183     \\
    \addlinespace[2pt]
    $R^{2}$              & 0.110       & 0.050       & 0.040        & 0.036       \\
    \bottomrule
  \end{tabular}
  \begin{tablenotes}
    \footnotesize
    \item \emph{Notes:} Cluster-robust standard errors at the regional level in parentheses. *** $p<0.01$, ** $p<0.05$, * $p<0.1$. Outcome variables (``Checkbook'', ``Total'') and the comprehension-based sample partition (``High/Low comprehension'') are defined as in Table~\ref{tab:compulsory}. Treatment effects are reported in JPY. At the time of the experiment, 1 USD was approximately 133 JPY. ITT estimates use random assignment to the opt-in condition as the regressor; ATE rows reproduce the compulsory estimates from Table~\ref{tab:compulsory} for comparison. ``Take-up rate'' is the share of opt-in participants who chose to receive the offered subsidy. All estimates use the budget-adjusted design (Section~\ref{sec:budget}). The matching advantage in the total amount received by the charity, statistically indistinguishable under compulsory assignment, re-emerges substantially under self-selection in Column (2).
  \end{tablenotes}
  \end{threeparttable}
\end{table}

\begin{table}[htbp]
  \centering
  \caption{Baseline Characteristics Predicting Take-up by Subsidy Scheme}
  \label{tab:takeup_regression}
  \begin{threeparttable}
  \small
  \setlength{\tabcolsep}{10pt}
  \renewcommand{\arraystretch}{1.15}
  \begin{tabular}{l c c}
    \toprule
    \emph{Linear Probability Model} & (1) & (2) \\
    Take-up for                     & 1:1 matching & 50\% rebate \\
    \midrule
    Baseline altruism        & 0.00022** & 0.00007   \\
                              & (0.00007) & (0.00010) \\
    \addlinespace[2pt]
    Female                   & 0.085     & 0.093     \\
                              & (0.101)   & (0.058)   \\
    \addlinespace[2pt]
    Age                      & 0.004     & $-0.002$ \\
                              & (0.004)   & (0.001)   \\
    \addlinespace[2pt]
    Married                  & $-0.098$     & 0.124  \\
                              & (0.062)   & (0.071)   \\
    \addlinespace[2pt]
    Having children          & 0.044  & $-0.177$**  \\
                              & (0.076)   & (0.078)   \\
    \addlinespace[2pt]
    Educational years        & 0.027*    & 0.011     \\
                              & (0.012)   & (0.010)   \\
    \addlinespace[2pt]
    Household income         & $-0.00002$  & $-0.00005$  \\
                              & (0.00007)   & (0.00012)   \\
    \addlinespace[2pt]
    No income information    & $-0.062$  & $-0.081$  \\
                              & (0.060)   & (0.085)   \\
    \addlinespace[2pt]
    Living in urban areas    & 0.113     & 0.064     \\
                              & (0.066)   & (0.047)   \\
    \addlinespace[6pt]
    Constant                 & $-0.074$  & 0.485**   \\
                              & (0.307)   & (0.190)   \\
    \midrule
    Observations             & 251       & 245       \\
    $R^{2}$                  & 0.069     & 0.040     \\
    \bottomrule
  \end{tabular}
  \begin{tablenotes}
    \footnotesize
    \item \emph{Notes:} Cluster-robust standard errors at the regional level are reported in parentheses. *** $p<0.01$, ** $p<0.05$, * $p<0.1$. Columns (1) and (2) report separate linear probability models for take-up of the 1:1 matching and 50\% rebate schemes, respectively, using participants assigned to the corresponding opt-in condition. The dependent variable equals one if the participant took up the offered scheme and zero otherwise. The analysis uses the high-comprehension sample, defined as participants scoring above the median on the eight-item matching-and-rebate comprehension check (six or more correct answers). Baseline altruism is measured in JPY, and household income in units of 10{,}000 JPY. At the time of the experiment, 1 USD was approximately 133 JPY.
  \end{tablenotes}
  \end{threeparttable}
\end{table}

\appendix
\thispagestyle{plain}



\setcounter{footnote}{0}
\setcounter{figure}{0}
\setcounter{table}{0}

\section{Theoretical Framework}
\label{app:theory}

\subsection{Setup of the \citet{hungerman2021impure} Model}
We set up the model. A donor derives utility from three components: the checkbook giving amount $g$, the total amount received by the charity $G$, and consumption $c$. The donor’s utility function is described as follows:
\begin{equation}
U(c,g,G)
=
c
+
\frac{\theta}{1+\frac{1}{e}}
\left(
\frac{g^{\gamma}G^{1-\gamma}}{\theta}
\right)^{1+\frac{1}{e}}
\label{eq:1}
\end{equation}
where $e$ is the price elasticity of donations $(e<0)$, $\theta$ is the altruistic preference parameter $(\theta \geq 0)$, and $\gamma$ is the warm-glow preference parameter $(0 \leq \gamma \leq 1)$.

The donor allocates both $g$ and $c$ with the following budget constraint:
\begin{equation}
c+g=y+tg
\label{eq:2}
\end{equation}
where $t$ is the rebate rate $(0 \leq t \leq 1)$. Under the rebate treatment, $tg$ is refunded to the donor when she/he chooses the checkbook giving amount $g$. In the control and the matching treatment with $t=0$, the budget constraint is $c+g=y$.

The relationship between the checkbook giving amount and the total amount received is $G=(1+m)g$, where $m$ is the matching rate $(0 \leq m)$. Under the $1{:}m$ matching treatment, the total amount received equals the donor’s checkbook giving amount $g$ plus $mg$ provided by a third party. In the control and the rebate treatment with $m=0$, this relationship is $G=g$\footnote{There is a difference between \citet{hungerman2021impure} model and ours with respect to the definition of $G$. In their model, $G$ explicitly includes donations made by others, and in the matching treatment group, the model is structured such that other people’s donations are also doubled. Since their empirical analysis uses non-experimental data, it is reasonable to assume that the model reflects such real-world complexities. By contrast, in our model, $G$ includes only the donor’s own contribution (and the doubled amount if applicable), based on the structure of our experimental design, and does not include others’ donations. The model assumes a simplified condition in which the total amount donated by others does not vary between the matching and rebate treatments.}.

Given the utility function and budget constraint, the optimal checkbook giving amount $g^*$ and the optimal total amount received $G^*$ are:
\begin{equation}
g^*(p_m,p_t)
=
\theta\, p_m^{(1+e)(1-\gamma)}\, p_t^{e}
\label{eq:3}
\end{equation}
\begin{equation}
G^*(p_m,p_t)
=
\theta\, p_m^{-\gamma + e(1-\gamma)}\, p_t^{e}
\label{eq:4}
\end{equation}
where $p_m=\frac{1}{1+m}$ and $p_t=1-t$ represent the decreases in donation prices caused by matching and rebate treatment, respectively. Hereafter, we call these the matching and rebate prices. Equations~(\ref{eq:3}) and~(\ref{eq:4}) show that the two optimal giving amounts are influenced by the prices as well as altruistic and warm-glow preferences. Notably, the warm-glow preference influences the checkbook and total amounts via the matching price, implying that the treatment effect of matching includes the warm-glow preference’s effect, whereas that of rebate does not.

We then examine the impact of changes in rebate and matching prices on checkbook and total amounts. We first check the elasticities of rebate and matching prices to prepare for this. The elasticity of the rebate price for the checkbook giving amount is:
\begin{equation}
e_r^g
=
\frac{\partial g^*(p_m,p_t)}{\partial p_t}
\frac{p_t}{g^*(p_m,p_t)}
=
e
\label{eq:5}
\end{equation}
The rebate price elasticity for checkbook giving equals the structural price elasticity $e$, with no role for warm-glow preferences. Similarly, the elasticity of the rebate price for the total amount received is:
\begin{equation}
e_r^G
=
\frac{\partial G^*(p_m,p_t)}{\partial p_t}
\frac{p_t}{G^*}
=
e
\label{eq:6}
\end{equation}
Both rebate elasticities equal $e$. They depend solely on the structural price elasticity, with no role for warm-glow.

Next, we examine the elasticity of the matching price for the checkbook giving amount:
\begin{equation}
e_m^g
=
\frac{\partial g^*(p_m,p_t)}{\partial p_m}
\frac{p_m}{g^*(p_m,p_t)}
=
(1+e)(1-\gamma)
\label{eq:7}
\end{equation}
This shows that, unlike the rebate, its price elasticity for the checkbook giving amount is influenced by warm-glow preference. The first term in Equation~(\ref{eq:7}) reduces the checkbook giving amount, while the warm-glow preference mitigates this decrease, as the donor values making their own contribution. The second term shows that the warm-glow preference weakens the donors’ price sensitivity. Similarly, the elasticity of the matching price for the total amount received is:
\begin{equation}
e_m^G
=
\frac{\partial G^*(p_m,p_t)}{\partial p_m}
\frac{p_m}{G^*(p_m,p_t)}
=
-\gamma + e(1-\gamma)
\label{eq:8}
\end{equation}
This elasticity is also influenced by warm-glow preferences.

We analyze the relationship in the price elasticities of matching between the checkbook giving amount and the total amount received. Equations~(\ref{eq:7}) and~(\ref{eq:8}) suggest that the relationship is $e_m^g = 1 + e_m^G$, implying that the price elasticity for the total amount received exceeds in absolute value that for the checkbook giving amount. This is due to strategic substitutability.

\subsection{Compulsory Treatment Effects: Derivation of Hypotheses 1 and 2}

Using the elasticities, we examine the treatment effects of rebates and matchings, compare the effects, and set hypotheses. The hypothesis for the checkbook giving amount is as follows:

\begin{quote}
\textbf{Hypothesis 1:} \emph{Regardless of warm-glow preferences, a decrease in the rebate price consistently leads to a larger increase in checkbook giving amounts than an equivalent decrease in the matching price.}
\end{quote}
If donors have no warm-glow preference $(\gamma=0)$, an increase in the checkbook giving amounts due to a drop in the rebate price $(e_r^g=e)$ is always larger than that due to a drop in the matching price $(e_m^g=1+e)$. This is strategic substitutability, meaning that third-party compensation reduces the checkbook giving amount, and it counteracts the increase in the checkbook giving amount caused by the decline in the matching price. Even if donors have warm-glow preferences $(\gamma>0)$, an increase in the checkbook giving amounts due to a drop in the rebate price $(e_r^g=e)$ is always larger than that due to a drop in the matching price $(e_m^g=(1+e)(1-\gamma))$.

The hypothesis for the total amount received is as follows:

\begin{quote}
\textbf{Hypothesis 2:} \emph{If donors have no warm-glow preference $(\gamma=0)$, the effects of changes in the rebate and matching prices on the total amount received by the charity are the same.}

\emph{If donors have warm-glow preferences $(\gamma>0)$ and are less responsive to donation prices $(-1<e<0)$, a decrease in the matching price leads to a larger increase in the total amount received than an equivalent decrease in the rebate price. If they are more responsive to donation prices $(e\leq -1)$, a decrease in the rebate price leads to a larger increase in the total amount received than an equivalent decrease in the matching price.}
\end{quote}
If donors have no warm-glow preference $(\gamma=0)$, the effects of matching and rebate prices on the total amount received are the same $(e_r^G=e_m^G=e)$.

If donors have warm-glow preferences $(\gamma>0)$, the relationship between the two treatment effects $(e_r^G=e,\ e_m^G=-\gamma+e(1-\gamma))$ depends on the original price elasticity of donations $(e)$. When donors are more responsive to donation prices $(e\leq -1)$, a drop in the rebate price leads to a greater increase in the total amount received than a drop in the matching price. This implies that the treatment effect of matching could be lower than that of the rebate when the impact of a drop in the matching price on the total amount received is heavily diminished by warm-glow preferences. Conversely, when donors are less responsive to donation prices $(-1<e<0)$, the increase in the total amount received from a drop in the matching price surpasses that with the rebate price drop. In this case, the increasing effect of warm-glow preferences on the total amount received is stronger, leading to the prediction that the treatment effect of matching exceeds that of the rebate.

The prior-literature averages of $e \approx -0.461$ and $\gamma \approx 0.511$ \citep{eckel2003rebate,eckel2006subsidizing,eckel2008subsidizing,eckel2017comparing,blumenthal2012subsidizing,scharf2015price,hungerman2021impure} satisfy $-1 < e < 0$ and $\gamma > 0$. Under these values, the model predicts that a drop in the matching price leads to a greater increase in the total amount received by the charity than a comparable decrease in the rebate price.

\subsection{Indirect Utility and Take-up: Derivation of Hypothesis 3}

Next, based on the impure impact model of \citet{hungerman2021impure}, we examine how the introduction of self-selection affects donors' take-up decisions and how the take-up rates may differ between the rebate and matching schemes. Using the indirect utility function, we show that the take-up rates depend on whether donors have warm-glow preferences.

We illustrate how a donor’s indirect utility changes depending on whether they choose to take up the rebate or the matching treatment. To do this, we first substitute Equations~(\ref{eq:3}) and~(\ref{eq:4}) into Equation~(\ref{eq:1}):
\begin{equation}
V(y,p_m,p_t)
=
y
-
\frac{p_t}{1+e}
g^*(p_m,p_t)
\label{eq:9}
\end{equation}

Donors will opt into a matching or rebate treatment when the utility gain from the price reduction exceeds the associated physical or psychological switching costs. The change in their indirect utility due to a change in the matching price or rebate price can be determined as follows:
\begin{equation}
\left.
\frac{\partial V}{\partial p_t}
\right|_{p_m=1}
=
-
G^*(1,p_t),
\qquad
\left.
\frac{\partial V}{\partial p_m}
\right|_{p_t=1}
=
-
(1-\gamma)
G^*(p_m,1)
\label{eq:10}
\end{equation}

According to Equation~(\ref{eq:4}), if donors have altruistic preferences $(\theta>0)$, their utility increases as both matching and rebate prices drop. It also increases monotonically in response to price reductions in both cases. Furthermore, for matching donations, if donors have warm-glow preferences $(\gamma>0)$, an increase in donors’ utility due to the price reduction drops in proportion to the degree of warm-glow.

In the realistic case where donors have altruistic preferences $(\theta>0)$ and face equivalent switching costs for opting into either scheme, the take-up decision depends on whether the utility gain from the price reduction exceeds the switching cost. Comparing the two schemes, when donors have no warm-glow preferences $(\gamma=0)$, the factor $(1-\gamma)=1$ in Equation~(\ref{eq:10}) makes the marginal utility gains from rebate and matching structurally equivalent, and the resulting total amount received by the charity $(G^*)$ is the same under both schemes. Hence, the two schemes yield equal take-up rates.

When donors have warm-glow preferences $(\gamma>0)$, the factor $(1-\gamma)<1$ dampens the matching scheme's utility gain, because a rebate lowers the effective prices of both warm-glow (checkbook giving) and impact $(G^*)$, while a matching lowers only the price of impact. Consequently, the utility gain from opting into the rebate exceeds that from the matching, and the take-up rate for rebate is predicted to be higher than that for matching.

\begin{quote}
\textbf{Hypothesis 3:} \emph{If donors have no warm-glow preferences $(\gamma=0)$, take-up rates are equal across the two schemes. If donors have warm-glow preferences $(\gamma>0)$, the take-up rate under the rebate scheme exceeds that under the matching scheme.}
\end{quote}

\subsection{Selection and Treatment Effects: Derivation of Hypothesis 4}

Equation~(\ref{eq:10}) also allows us to derive predictions for how self-selection shapes treatment effects. We now assume that donors are heterogeneous in their altruistic preference $\theta$ across individuals. 
Based on Equation~(\ref{eq:10}), the gain in indirect utility from introducing either the rebate or matching scheme is proportional to the total amount received $G^*$ at the treated price. Since $G^*$ scales linearly with $\theta$ (see Equations~(\ref{eq:3}) and~(\ref{eq:4})), donors with larger $\theta$ obtain larger utility gains from opting into either scheme.
Under the same switching cost as assumed in Section~A.3, donors take up the intervention when their utility gain exceeds the switching cost, which translates into a threshold rule on $\theta$: donors with sufficiently large $\theta$ opt in. Since the treatment effect on the total amount received also scales linearly with $\theta$, those who self-select into the treatment are precisely those with the largest treatment effects.
As a result, the treatment-on-the-treated (TOT) exceeds the average treatment effect (ATE) for both schemes.

\begin{quote}
\textbf{Hypothesis 4:} \emph{For both schemes, individuals who choose to take up the intervention are expected to experience larger increases in total amounts received by the charity than those assigned compulsorily; that is, $\mathrm{TOT} > \mathrm{ATE}$.}
\end{quote}

\newpage


\section{Survey Items and Experimental Interface}
\label{app:survey}

\noindent\textit{Note on terminology: The experimental instructions used the term ``tax deduction,'' the real-world policy analog of the rebate scheme, in describing the incentive. The underlying mechanism is identical to the 50\% rebate described in the main text.}

\subsection{Screening Survey}

\subsubsection{Screening Survey Items}

\noindent\textit{The screening survey was conducted between February 10 and 15, 2023, by Sasaki Lab at The University of Osaka via MyVoice.com Ltd. Items appear in the order presented to respondents.}

\bigskip

\noindent
\textit{Item index:}
S1--S3 demographics $\cdot$ S4 hypothetical donation (baseline altruism) $\cdot$ S5 knowledge of donation schemes $\cdot$ S6--S9 comprehension check (8 sub-questions; matching/rebate, randomized pair order) $\cdot$ S10--S11 willingness to pay $\cdot$ S12 nature connectedness scale.

\bigskip
\hrule
\bigskip

\begin{center}
    \textbf{Background and donation experience (I0--S5)}
\end{center}

\paragraph{I0. Informed consent}

\paragraph{S1. Sex} \quad

\noindent\textbf{Question:} ``What is your sex?''

\noindent\textbf{Response:} single choice --- Male / Female / Other.

\paragraph{S2. Age} \quad

\noindent\textbf{Question:} ``What is your age?''

\noindent\textbf{Response:} numeric entry in years.

\paragraph{S3. Residential area} \quad

\noindent\textbf{Question:} ``Where do you currently reside?''

\noindent\textbf{Response:} single choice; list of 47 Japanese prefectures plus ``Other.''

\paragraph{S4. Hypothetical donation amount (baseline altruism)} \quad

\noindent\textbf{Question:} ``Imagine you have unexpectedly received 1,000 JPY today (ten 100-JPY coins). If a donation box for an activity to protect Japan’s fragile natural environment were placed in front of you, how much of the 1,000 JPY would you be willing to donate? Please answer in 100-JPY increments. Note: Any amount you keep can be spent on anything you choose other than donation.''

\noindent\textbf{Response:} single choice from 0, 100, 200, \ldots, 1,000 JPY. After selection, respondents pressed a ``Calculate amount'' button to view (i) the checkbook amount, (ii) the reward to self, and (iii) the total amount received by the charity.

\paragraph{S5. Knowledge and experience of donation schemes} \quad

\noindent\textbf{Question:} ``Are you aware of, or have you ever used, the following schemes for monetary donations?''
Two schemes were described:
\begin{itemize}
    \item[-] \textit{Tax deduction:} ``A scheme that reduces income tax and resident tax under certain conditions when individuals donate money.''  
    \item[-] \textit{Matching gift:} ``A scheme in which a company, foundation, or similar entity adds a fixed proportion to an individual’s donation.''
\end{itemize}

\noindent\textbf{Response:} single choice per system --- ``I have used it before'' / ``I know about it but have never used it'' / ``I did not know.''

\bigskip
\hrule
\bigskip

\begin{center}
    \textbf{Comprehension checks on matching gift and rebate schemes (S6--S9)}
\end{center}

\noindent\textit{Note: Items S6--S9 form the eight-question comprehension check (each item contains two sub-questions, the reward to self and the total amount received by the charity). They are presented as two fixed-order pairs: S6--S7 (matching gift, 400 JPY then 800 JPY) and S8--S9 (tax deduction / rebate, 400 JPY then 800 JPY). The presentation order of the two pairs was randomized across respondents, so that some respondents saw the matching pair first and others saw the rebate pair first.}

\paragraph{S6. Comprehension check --- Matching gift (400 JPY donation)} \quad

\noindent\textbf{Scenario:} respondents were asked to imagine donating part of 1,000 JPY using a matching gift, described as:
``A third party adds an amount equal to your donation and delivers it to the charity.''
Two worked examples were provided (donating 1,000 JPY $\rightarrow$ reward to self 0 JPY, total amount received 2,000 JPY; donating 0 JPY $\rightarrow$ reward to self 1,000 JPY, total amount received 0 JPY).

\noindent\textbf{Question:} ``If you donate 400 JPY under this scheme, what would be (a) the reward to self, and (b) the total amount received by the charity? Please answer in 100 JPY increments.''

\noindent\textbf{Response:} single choice per outcome from 0, 100, \ldots, 2,000 JPY. Two sub-questions counted as two items of the comprehension check.

\noindent\textbf{Correct answers (shown to respondents after answering):} 600 JPY (reward to self); 800 JPY (total amount received by the charity).

\paragraph{S7. Comprehension check --- Matching gift (800 JPY donation)} \quad

\noindent Same scenario as S6, but for an 800 JPY donation. Two sub-questions.

\noindent\textbf{Correct answers:} 200 JPY (reward to self); 1,600 JPY (total amount received by the charity).

\paragraph{S8. Comprehension check --- Tax deduction / rebate (400 JPY donation)} \quad

\noindent\textbf{Scenario:} respondents were asked to imagine donating part of 1,000 JPY using the tax deduction scheme, described as: ``A third party covers 50\% of your donation amount and refunds it to you.''
Two worked examples were provided: (donating 1,000 JPY $\rightarrow$ reward to self 500 JPY, total amount received 1,000 JPY; donating 0 JPY $\rightarrow$ reward to self 1,000 JPY, total amount received 0 JPY).

\noindent\textbf{Question:} ``If you donate 400 JPY under this scheme, what would be (a) the reward to self, and (b) the total amount received by the charity? Please answer in 100 JPY increments.''

\noindent\textbf{Response:} single choice per outcome from 0, 100, \ldots, 2,000 JPY. Two sub-questions.

\noindent\textbf{Correct answers:} 800 JPY (reward to self; $=1,000-400+200$ refund); 400 JPY (total amount received by the charity).

\paragraph{S9. Comprehension check --- Tax deduction / rebate (800 JPY donation)} \quad

\noindent Same scenario as S8, but for an 800 JPY donation. Two sub-questions.

\noindent\textbf{Correct answers:} 600 JPY (reward to self; $=1,000-800+400$ refund); 800 JPY (total amount received by the charity).

\bigskip
\hrule
\bigskip

\begin{center}
    \textbf{Willingness to pay and nature connectedness scale (S10--S12)}
\end{center}

\paragraph{S10. Willingness to pay for matching gift (by fee level)} \quad

\noindent\textbf{Scenario:} ``Imagine donating part of 1{,}000 JPY to protect Japan’s fragile natural environment. To use the matching gift, you must pay a `scheme usage fee' separately from the donation.''

\noindent\textbf{Matching gift definition (repeated):} ``A third party adds an amount equal to your donation and delivers it to the charity.''

\noindent\textbf{Question:} ``Would you use the matching gift at each of the following fee levels?''

\noindent\textbf{Response:} per fee level --- Use / Do not use. 
Fee levels: 0, 10, 50, 100, 150, 200, 250, 300 JPY.

\paragraph{S11. Willingness to pay for tax deduction / rebate (by fee level)} \quad

\noindent Same as S10, but for the tax deduction (50\% rebate) scheme.
Fee levels and response format: identical.

\paragraph{S12. Nature connectedness scale (6 items)} \quad

\noindent\textbf{Question:} ``To what extent do each of the following statements apply to you? Please select on a 5-point scale, where 5 = `Applies very much' and 1 = `Does not apply at all.'''

\noindent\textbf{Response:} 5-point Likert per item --- 5: Applies very much / 4: Somewhat applies / 3: Neither / 2: Somewhat does not apply / 1: Does not apply at all.

\noindent\textbf{Items:}
\begin{enumerate}
    \item The ideal place for me to spend a vacation is a remote area surrounded by abundant nature.
    
    \item I am someone who is always thinking about how my actions affect nature.
    
    \item Interacting with nature serves as my spiritual support (mental anchor).
    
    \item I am someone who always pays attention to wildlife.
    
    \item Interacting with nature is important for being myself.
    
    \item I feel a strong connection to the earth and all living things on it.
\end{enumerate}

\newpage

\subsubsection{Selected Screenshots from Screening Survey}

\begin{figure}[htbp]
  \centering
  \includegraphics[width=0.8\linewidth]{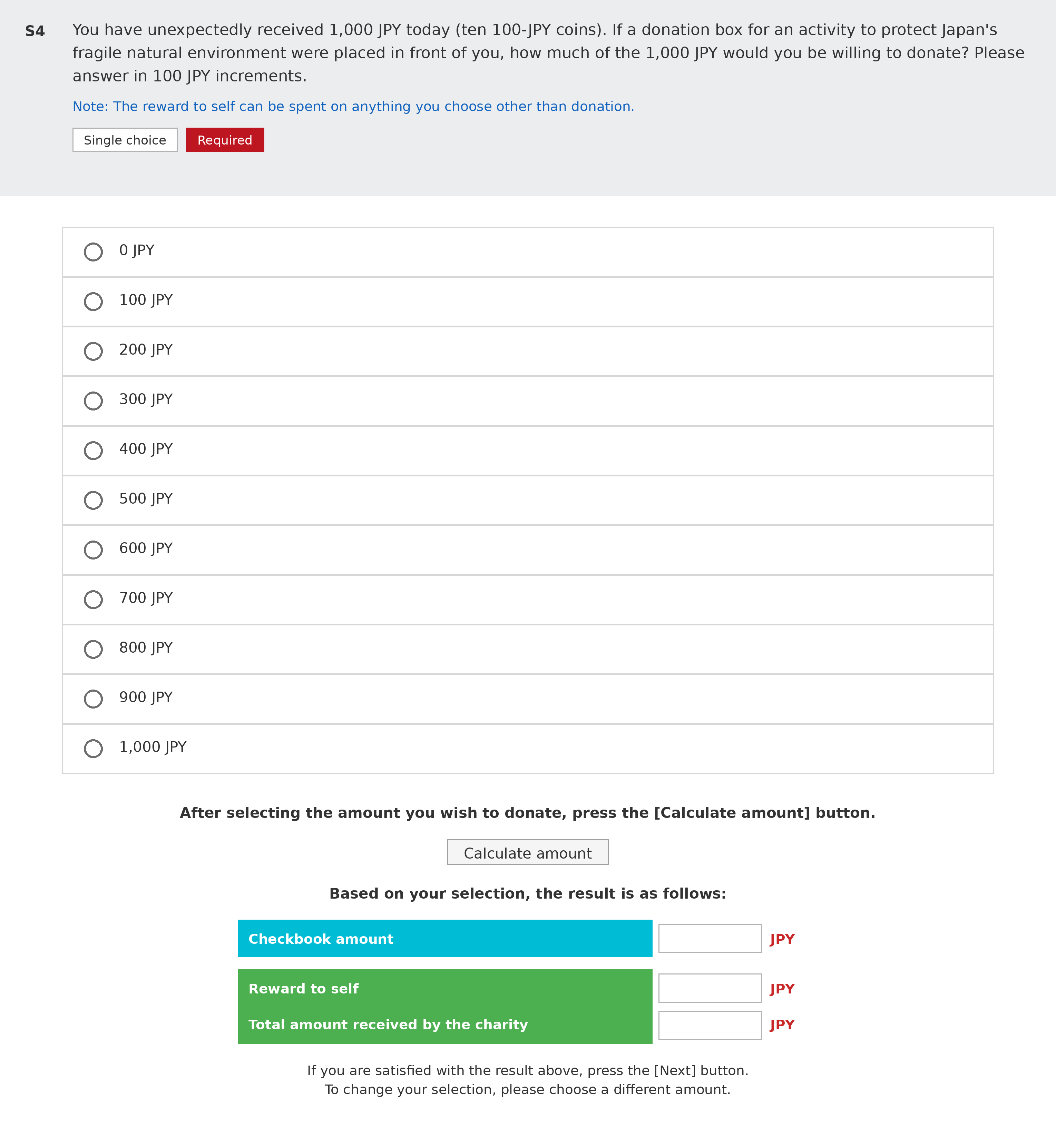}
  \caption{S4: Hypothetical donation amount question.}
  \label{fig:s4_donation}
\end{figure}

\begin{figure}[htbp]
  \centering
  \includegraphics[width=0.8\linewidth]{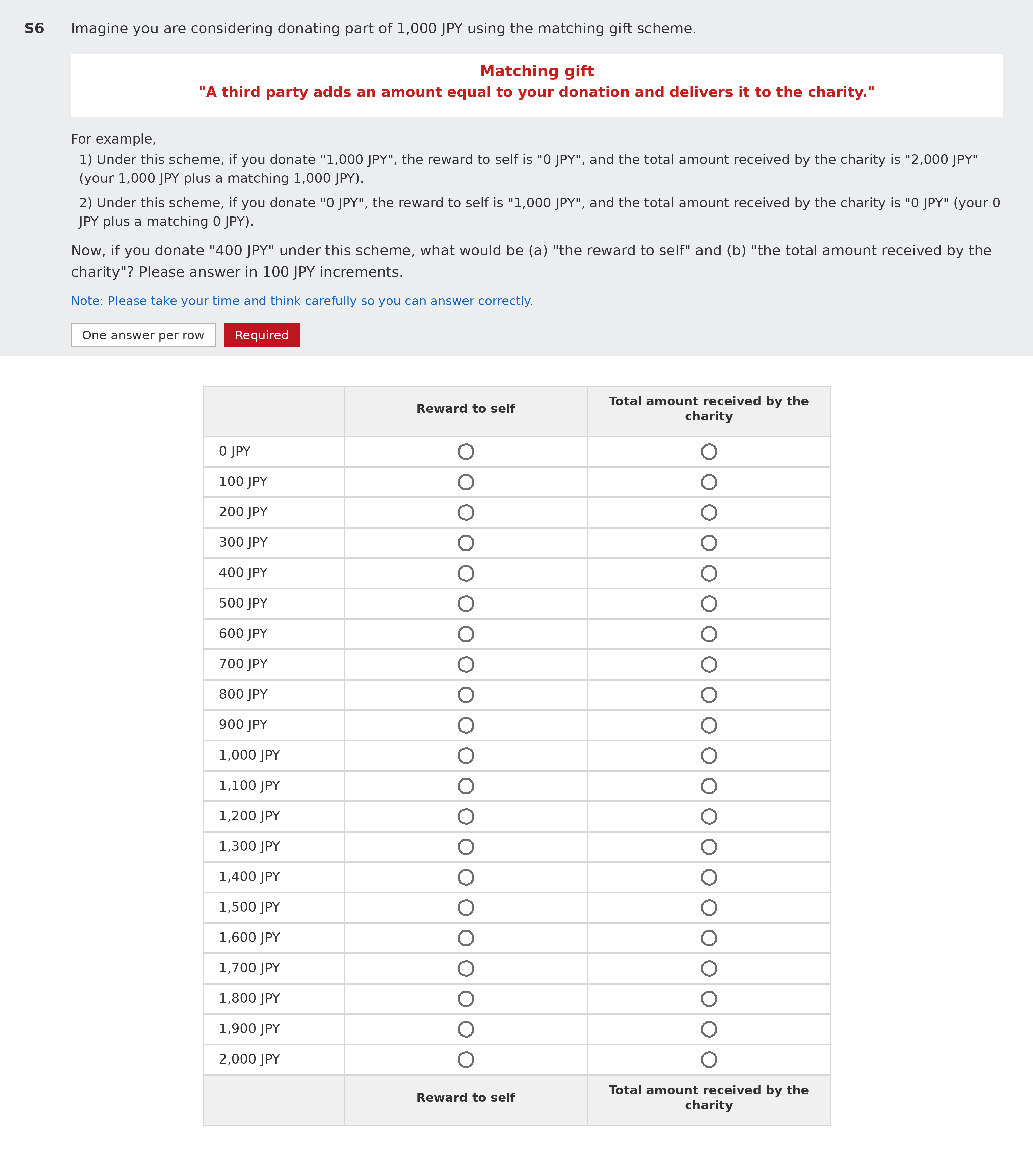}
  \caption{S6: Comprehension check on the matching scheme (400 JPY donation).}
  \label{fig:s6_match}
\end{figure}

\begin{figure}[htbp]
  \centering
  \includegraphics[width=0.8\linewidth]{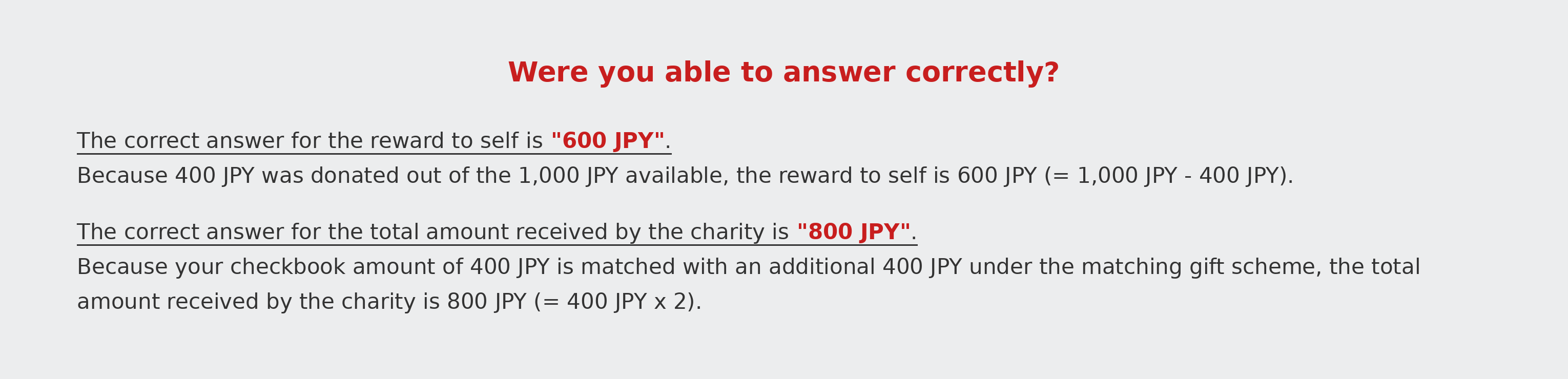}
  \caption{S6 feedback: Correct-answer display.}
  \label{fig:s6_feedback}
\end{figure}

\begin{figure}[htbp]
  \centering
  \includegraphics[width=0.8\linewidth]{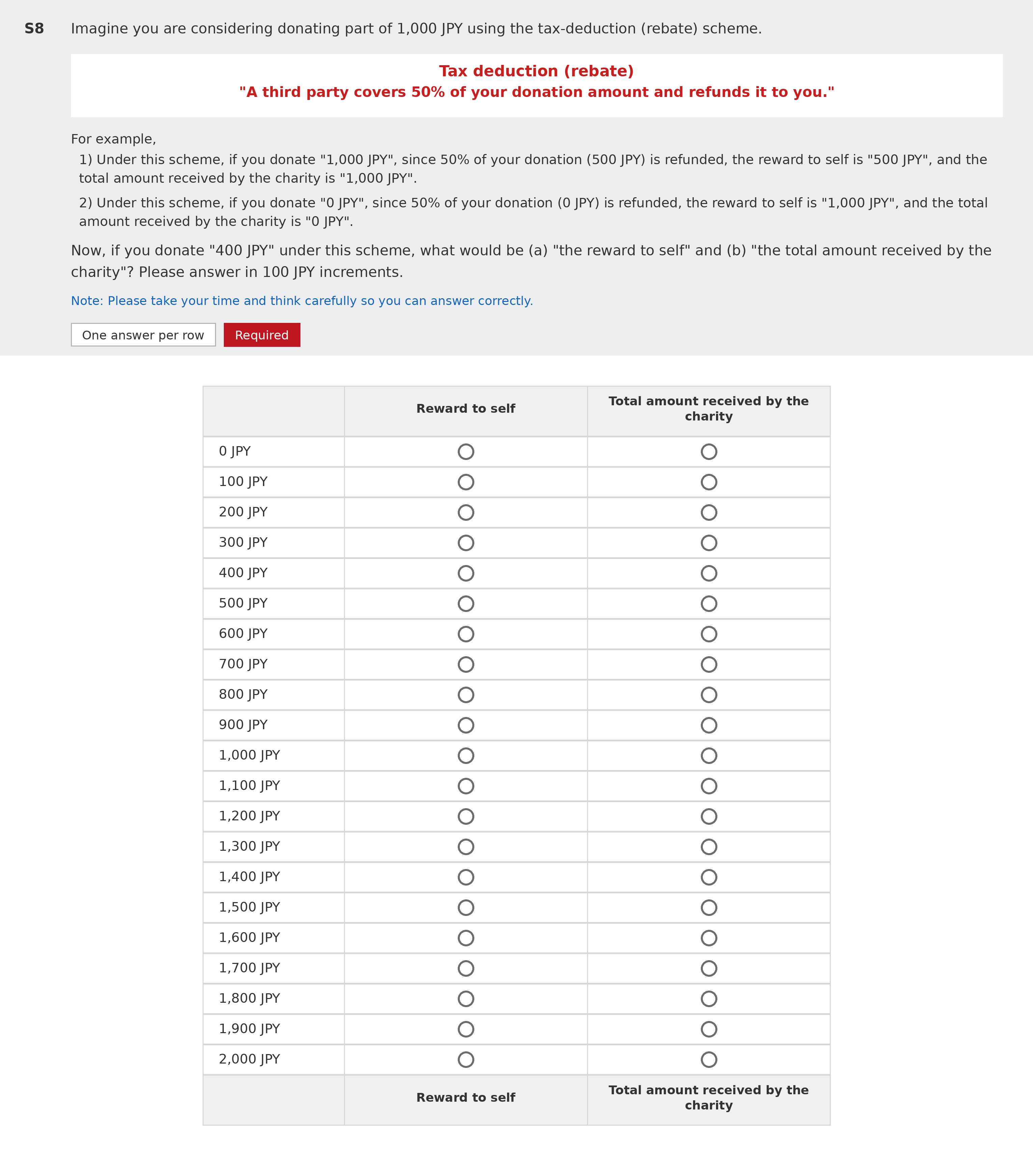}
  \caption{S8: Comprehension check on the rebate scheme (400 JPY donation).}
  \label{fig:s8_rebate}
\end{figure}

\begin{figure}[htbp]
  \centering
  \includegraphics[width=0.8\linewidth]{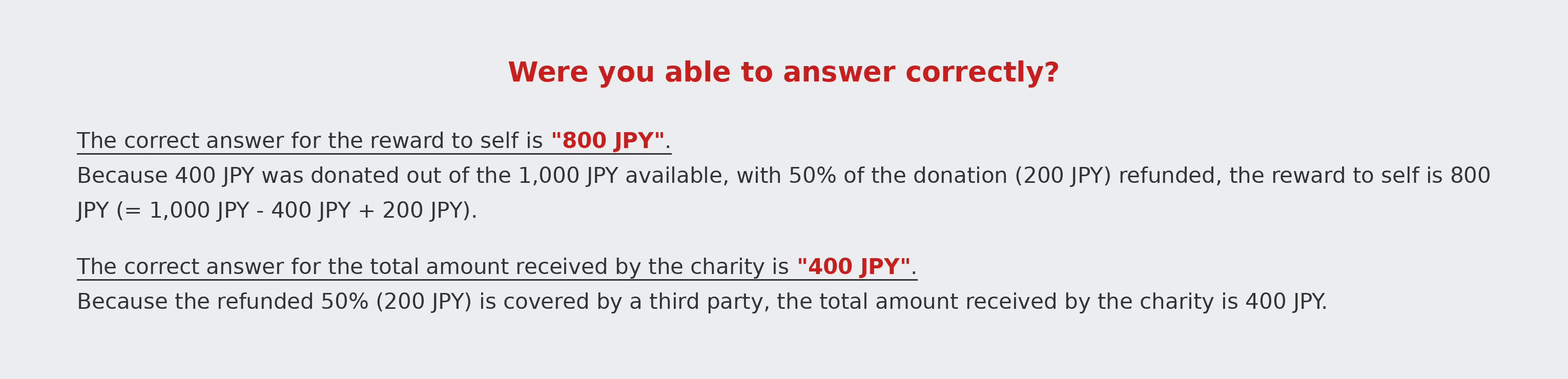}
  \caption{S8 feedback: Correct-answer display.}
  \label{fig:s8_feedback}
\end{figure}

\newpage

\subsection{Main Experiment}

\subsubsection{Main Experiment Items}

\noindent
\textit{The main experiment was conducted between February 17 and 21, 2023, by Sasaki Lab at The University of Osaka via MyVoice.com Ltd. Only respondents who completed the screening survey (B.1.1) were invited. Items appear in the order presented to respondents.}

\bigskip

\noindent
\textit{Item index:}
Q1 behavioral attitudes (15 items incl.\ 1 attention check)
$\cdot$
Q2--Q5 preference elicitation (patience, risk, lottery, self-esteem)
$\cdot$
Q6--Q8 prior donation experience
$\cdot$
Q8SA--Q8S5C incentivized donation experiment
$\cdot$
Q9--Q10 post-decision self-reports
$\cdot$
Q11--Q12 hypothetical donation experiment
$\cdot$
Q13--Q19 demographics and socioeconomic characteristics.

\paragraph{I0. Informed Consent} \quad

\bigskip
\hrule
\bigskip

\begin{center}
    \textbf{Behavioral and preference measures (Q1–Q5)}
\end{center}

\paragraph{Q1. Behavioral and pro-social attitudes (15 items, including 1 attention check)} \quad

\noindent\textbf{Question:} ``To what extent do each of the following statements apply to you? Please select on a 5-point scale, where 5 = `Applies very much' and 1 = `Does not apply at all.'''

\noindent\textbf{Response:} 5-point Likert per item ---
5: Applies very much /
4: Somewhat applies /
3: Neither /
2: Somewhat does not apply /
1: Does not apply at all.

\noindent\textbf{Items:}
\begin{enumerate}
    \item I feel happy when I see others happy.
    
    \item I am very concerned about how others evaluate me.
    
    \item I am someone who takes the initiative on tasks others would rather not do.
    
    \item If I had enough money to spare, I would consider donating.
    
    \item I would want to donate even if it meant cutting back on money I could spend on myself.
    
    \item When someone does me a favor, I want to repay them.
    
    \item I would want to retaliate against someone who treats me unfairly, even if it is costly.
    
    \item I feel reassured when I behave like those around me.
    
    \item I do not cut in line.
    
    \item I willingly cooperate when doing something with others.
    
    \item I enjoy accomplishing something through cooperation with others.
    
    \item I commit to team play rather than individual play.
    
    \item I believe solving societal problems is the government's responsibility.
    
    \item I pay a large amount of tax.
    
    \item {[}Attention check{]} For this question, please select the leftmost option.
\end{enumerate}

\noindent\textit{Note: An on-screen warning was displayed on the next page to respondents who failed the attention check item (15), asking them to read questions more carefully.}

\paragraph{Q2. Patience (time preference)} \quad

\noindent\textbf{Question:} ``Generally speaking, how willing are you to give up something that benefits you now in order to gain more in the future? Please choose the option that best applies on an 11-point scale, where 10 = `Very willing to give up' and 0 = `Not at all willing to give up.'''

\noindent\textbf{Response:} single choice on an 11-point scale (0--10).

\paragraph{Q3. Risk tolerance} \quad

\noindent\textbf{Question:} ``Generally speaking, how willing or unwilling are you to take risks? Please choose the option that best applies on an 11-point scale, where 10 = `Very willing to take risks' and 0 = `Not at all willing to take risks.'''

\noindent\textbf{Response:} single choice on an 11-point scale (0--10).

\paragraph{Q4. Lottery choices (7 rows)} \quad

\noindent\textbf{Question:} ``Suppose you are choosing how to receive your monthly salary. Which of the following options, A or B, would be more desirable to you? Both options offer the same job content. (For students, homemakers, or non-earners, please assume your income equals your monthly expenditure.) For each of the seven rows, please choose the option you prefer.''

\noindent\textbf{Response:} single choice per row (A or B).

\noindent\textbf{Column entries (table header):}
\begin{itemize}
    \item[-] Option A: a 50/50 lottery whose loss size varies by row (described below).
    \item[-] Option B: a certain 0.5\% increase (the same text in all seven rows).
\end{itemize}

\noindent\textbf{Row entries (table side; Option A varies across rows):}
\begin{enumerate}
    \item 50\% chance to double, 50\% chance to decrease by 60\%
    \item 50\% chance to double, 50\% chance to decrease by 50\%    
    \item 50\% chance to double, 50\% chance to decrease by 45\%    
    \item 50\% chance to double, 50\% chance to decrease by 30\%    
    \item 50\% chance to double, 50\% chance to decrease by 10\%    
    \item 50\% chance to double, 50\% chance to decrease by 5\%    
    \item 50\% chance to double, 50\% chance to decrease by 1\%
\end{enumerate}

\paragraph{Q5. Rosenberg Self-Esteem Scale (10 items)} \quad

\noindent\textbf{Question:} ``For each of the following characteristics, please indicate how much it applies to you. Please answer based on how you see yourself, not how you think others see you.''

\noindent\textbf{Response:} 5-point Likert per item ---
Applies /
Somewhat applies /
Neither /
Somewhat does not apply /
Does not apply.

\noindent\textbf{Items:}
\begin{enumerate}
    \item I feel that I am a person of worth, at least on an equal plane with others.    
    \item I feel that I have a number of good qualities.
    \item All in all, I am inclined to feel that I am a failure.
    \item I am able to do things as well as most other people.
    \item I feel I do not have much to be proud of.    
    \item I take a positive attitude toward myself.    
    \item On the whole, I am satisfied with myself.    
    \item I wish I could have more respect for myself.
    \item I certainly feel useless at times.    
    \item At times I think I am no good at all.
\end{enumerate}

\bigskip
\hrule
\bigskip

\begin{center}
    \textbf{Prior donation experience (Q6–Q8)}
\end{center}

\paragraph{Q6. Annual donation amount in 2022} \quad

\noindent\textbf{Question:} ``Approximately how much did you donate in monetary donations during the year 2022 (January--December)?''

\noindent\textbf{Notes provided to respondents:} --``Monetary donation'' refers to voluntarily contributing money to public-interest activities or organizations (not for yourself or your family). --It includes general donations as well as donations made via credit-card or mileage points, charity-linked product purchases, hometown tax (\textit{furusato nōzei}), and crowdfunding support.

\noindent\textbf{Response:} single choice. Categories: 0 JPY / 1--under 500 / 500--under 1{,}000 / 1{,}000--under 3{,}000 / 3{,}000--under 5{,}000 / 5{,}000--under 10{,}000 / 10{,}000--under 25{,}000 / 25{,}000--under 50{,}000 / 50{,}000--under 100{,}000 / 100{,}000--under 500{,}000 / 500{,}000--under 1{,}000{,}000 / 1{,}000{,}000 JPY or more.

\paragraph{Q7. Interest in monetary donation} \quad

\noindent\textbf{Question:} ``How interested are you in making monetary donations?''

\noindent\textbf{Response:} single choice ---
Interested /
Somewhat interested /
Somewhat not interested /
Not interested.

\paragraph{Q8. Categories of donations made in 2022 (16 categories)} \quad

\noindent\textbf{Question:} ``To which of the following organizations or activities did you make monetary donations during 2022 (January--December)?''

\noindent\textbf{Response:} per category ---
A (donated) /
B (did not donate).

\noindent\textbf{Categories:} (1) COVID-19 related organizations or activities; (2) community development; (3) emergency disaster relief; (4) international cooperation/exchange; (5) arts, culture, and sports; (6) education and research; (7) employment promotion/support; (8) health, medical care, and welfare; (9) child and youth development; (10) nature and environmental conservation; (11) rights advocacy and support; (12) intermediary support for social-contribution activities; (13) hometown tax (return-gift-oriented); (14) hometown tax (non-return-gift-oriented); (15) national or local government (other than hometown tax); (16) organizations/activities other than 1--15.

\bigskip
\hrule
\bigskip

\begin{center}
    \textbf{Incentivized donation experiment (Q8SA–Q8S5C)}
\end{center}

\noindent\textit{Note: The items above (Q1--Q8) elicited baseline behavioral characteristics and prior donation experience. From Q8SA onward, respondents entered the financially incentivized donation experiment, which provides the primary outcome variables used in the main text. The experiment proceeded as follows:
lottery information and comprehension check (Q8SA)
$\rightarrow$
charity introduction (I8S)
$\rightarrow$
treatment instructions for the assigned scheme (I8S2 / I8S4; matching or rebate)
$\rightarrow$
opt-in decision for opt-in groups (Q8S3O or Q8S5O)
$\rightarrow$
donation decisions under two budget caps (Q8S[1--5]B and Q8S[1--5]C).}

\paragraph{Q8SA. Lottery information and comprehension checklist (5 checkboxes)} \quad

\noindent\textbf{Respondents were informed:} -- There is a 1-in-10 chance to win an additional 1{,}000 JPY in reward points; winners will be notified by email. -- On the next page, a social-contribution project will be introduced. Respondents may donate part or all of the 1{,}000 JPY to the project. -- Assuming they win the 1{,}000 JPY, respondents will be asked twice how much they would donate. If they win, one of the two answers will be randomly selected and the donation will be executed accordingly.

\noindent\textbf{Response:} respondents had to check all of the following 5 items (all required) before proceeding:
\begin{enumerate}
    \item There is a real 1-in-10 chance of winning the additional 1{,}000 JPY in reward points.
    \item If you win, one of your two answers is randomly selected and the donation will be executed as you answered (the remaining amount is yours to keep as additional reward).    
    \item You cannot change your answers after winning, so please consider them carefully.
    \item If you do not win, you receive nothing additional and no donation is executed.    
    \item Winners will be notified by email at a later date.
\end{enumerate}

\paragraph{I8S. Charity introduction (with images)} \quad

\noindent A two-image introduction page showed the recipient organization and its mission. Images depicted
(1) the Karibayama--Ōhirayama forest ecosystem protection area in Hokkaido and
(2) the Chiribishi coral colony in Ōura Bay, Nago, Okinawa.

\noindent\textbf{Text presented:}
``Biological diversity refers to the connections in which diverse living things rely on one another. Earth is said to host millions of species, but many are at risk of extinction. Your donation will support `activities to protect Japan's fragile natural environment.' The recipient is the Nature Conservation Society of Japan, a public-interest incorporated foundation.''

\paragraph{I8S2 / I8S4. Treatment instructions (varies by group)} \quad

\noindent Two variants were displayed depending on group assignment:
\begin{itemize}
    \item[-] \textit{Matching variant (Groups 2 and 3):}
    ``A matching gift scheme has been introduced. Our research team will add an amount equal to your donation and deliver it to the charity.''    
    Four worked examples were shown for donations of 1{,}000, 800, 400, and 0 JPY (illustrating the reward to self and the total amount received by the charity).    
    \item[-] \textit{Rebate variant (Groups 4 and 5):}
    ``A tax-deduction (rebate) scheme has been introduced. Our research team will cover 50\% of your donation amount and refund it to you.''    
    Four worked examples were shown for the same donation levels.
\end{itemize}

\paragraph{Q8S3O. Opt-in decision --- Matching (Group 3)} \quad

\noindent After viewing the matching instructions, Group 3 respondents were asked:

\noindent\textbf{Question:}
``You may choose whether or not to use this matching gift scheme. To consider your donation under this scheme, select `Use' and proceed. To consider your donation without this scheme, leave `Do not use' selected and proceed.''

\noindent\textbf{Response:} single choice ---
Use / Do not use (default).

\paragraph{Q8S5O. Opt-in decision --- Rebate (Group 5)} \quad

\noindent After viewing the rebate instructions, Group 5 respondents were asked the parallel question for the rebate scheme.

\noindent\textbf{Response:} single choice ---
Use / Do not use (default).

\paragraph{Q8S1B--Q8S5C. Main donation decision (per group $\times$ two budget caps)} \quad

\noindent The donation decision was administered separately to each of the five groups, in two variants distinguished by the upper limit on the initial donation amount: a 1{,}000 JPY cap (variant B) and a 500 JPY cap (variant C). The order in which the two caps were presented (1{,}000-cap first or 500-cap first) was randomized across respondents.

\noindent\textbf{Common question stem:}
``If you win 1{,}000 JPY in the lottery, you can donate up to [1{,}000 JPY / 500 JPY] to the activity to protect Japan’s fragile natural environment. How much are you willing to donate? Please answer in 100 JPY increments.''

\noindent\textbf{Notes provided:}
-- ``The reward points remaining for you will be delivered later and can be exchanged for Amazon gift cards, e-gifts, book cards, PeX, etc.''
-- (For matching groups)
 ``Our research team will add an amount equal to your donation and deliver it to the charity.''
-- (For rebate groups)
 ``Our research team will cover 50\% of your donation amount and refund it to you.''

\noindent\textbf{Response:} single choice from 0, 100, 200, \ldots, 1{,}000 JPY. After selection, respondents pressed a ``Calculate amount'' button to view
(i) the checkbook amount,
(ii) the reward to self, and
(iii) the total amount received by the charity.

\noindent\textit{Note: Group-by-group identifiers
(B = 1{,}000 JPY cap, C = 500 JPY cap):
\begin{enumerate}
    \item Control --- Q8S1B, Q8S1C.
    \item Matching (compulsory) --- Q8S2B, Q8S2C.
    \item Matching (opt-in, takers) --- Q8S3B, Q8S3C.
    \item Rebate (compulsory) --- Q8S4B, Q8S4C.
    \item Rebate (opt-in, takers) --- Q8S5B, Q8S5C.
\end{enumerate}
Respondents in Groups 3 and 5 who selected ``Do not use'' in the opt-in decision proceeded under the same control-condition donation decision.}

\bigskip
\hrule
\bigskip

\begin{center}
    \textbf{Post-decision self-reports (Q9–Q10)}
\end{center}

\paragraph{Q9S1. Reference weight self-report --- first donation question} \quad

\noindent\textbf{Question:} ``When you chose your donation amount in the first of the two donation questions, how much did you refer to the automatically calculated `reward to self' versus `total amount received by the charity'? Please answer in 10\% increments so that the two sum to 100\%.''

\noindent\textbf{Response:} single choice from 11 options ranging from
``reward to self = 0\%; total amount received by the charity = 100\%''
to
``reward to self = 100\%; total amount received by the charity = 0\%.''

\paragraph{Q9S2. Reference weight self-report --- second donation question} \quad

\noindent Same question as Q9S1, but for the second donation question.

\paragraph{Q10. Trust and comprehension self-assessment (4 items)} \quad

\noindent\textbf{Question:} ``For each of the following statements about the two donation questions, how much do you agree? Please select on a 5-point scale, where 5 = `Strongly agree' and 1 = `Strongly disagree.'''

\noindent\textbf{Response:} 5-point Likert per item ---
5: Strongly agree /
4: Somewhat agree /
3: Neither /
2: Somewhat disagree /
1: Strongly disagree.

\noindent\textbf{Items:}
\begin{enumerate}
    \item I think the recipient ``Nature Conservation Society of Japan'' is trustworthy.
    \item I carefully confirmed the automatically calculated ``reward to self'' and ``total amount received by the charity'' and am satisfied with my decision.    
    \item I believe I will definitely receive the 1{,}000 JPY.
    \item If I win the 1{,}000 JPY, I believe the responsible researcher will deliver the amount I selected to the recipient.
\end{enumerate}

\bigskip
\hrule
\bigskip

\begin{center}
    \textbf{Hypothetical donation experiment \\ (Q11--Q12 and variants; presentation order randomized)}
\end{center}

\paragraph{Q11 / Q12. Hypothetical donation amount, no scheme (Control)} \quad

\noindent\textbf{Question:} ``Today, you have unexpectedly received 10{,}000 JPY. You may donate up to [10{,}000 JPY / 5{,}000 JPY] to the activity to protect Japan’s fragile natural environment. How much are you willing to donate? Please choose the option closest to your desired amount.''

\noindent\textbf{Response:} single choice from 20 options under the 10{,}000 JPY cap or 15 options under the 5{,}000 JPY cap:
0, 100, 200, \ldots, 1{,}000 JPY,
then bracketed ranges
1{,}100--2{,}000,
2{,}100--3{,}000,
\ldots,
9{,}100--10{,}000 JPY.
\begin{itemize}
    \item[-] Q11: 10{,}000 JPY upper limit.
    \item[-] Q12: 5{,}000 JPY upper limit.
\end{itemize}

\paragraph{Q11S1 / Q12S1. Hypothetical donation amount, matching gift (Matching treated)} \quad

\noindent Same question stem as Q11/Q12, with an additional instruction:
``Under the following scheme, please choose the option closest to your desired amount.
\textit{Matching gift: `A third party adds an amount equal to your donation and delivers it to the charity.'}''

\noindent\textbf{Response:} same choice set as Q11/Q12.
\begin{itemize}
    \item[-] Q11S1: 10{,}000 JPY upper limit.    
    \item[-] Q12S1: 5{,}000 JPY upper limit.
\end{itemize}

\paragraph{Q11S2 / Q12S2. Hypothetical donation amount, rebate (Rebate treated)} \quad

\noindent Same question stem as Q11/Q12, with an additional instruction:
``Under the following scheme, please choose the option that applies.
Tax deduction (rebate): `A third party covers 50\% of your donation amount and refunds it to you.'''

\noindent\textbf{Response:} same choice set as Q11/Q12.
\begin{itemize}
    \item[-] Q11S2: 10{,}000 JPY upper limit.    
    \item[-] Q12S2: 5{,}000 JPY upper limit.
\end{itemize}

\bigskip
\hrule
\bigskip

\begin{center}
    \textbf{Demographics and socioeconomic characteristics (Q13–Q19)}
\end{center}

\paragraph{Q13. Residential city size} \quad

\noindent\textbf{Question:} ``What is the size of the municipality where you currently live?''

\noindent\textbf{Response:} single choice ---
Government-designated city or one of the 23 Tokyo wards /
City with population of 300{,}000 or more /
City with population of 100{,}000--300{,}000 /
City with population under 100{,}000 /
Town or village.

\paragraph{Q14. Final school attended (highest education)} \quad

\noindent\textbf{Question:} ``Please indicate the last school you attended (or are currently attending).''

\noindent\textbf{Response:} single choice ---
Junior high school (graduated) /
High school (left without completion) /
High school (graduated) /
Junior college (left without completion; includes technical college) /
Junior college (graduated; includes technical college) /
University (left without completion) /
University (graduated) /
Master's program (left without completion) /
Master's program (completed) /
Doctoral program (left without completion) /
Doctoral program (completed).

\paragraph{Q15. Marital status} \quad

\noindent\textbf{Question:} ``Please indicate the option that applies to your spouse. Common-law spouses are included.''

\noindent\textbf{Response:} single choice ---
Currently have a spouse (husband or wife; including common-law) /
Separated /
Widowed /
Never married.

\paragraph{Q16. Household composition (5 items)} \quad

\noindent\textbf{Question:} ``Does your household currently include any of the following?''

\noindent\textbf{Response:} per item --- Yes / No.

\noindent\textbf{Items:} (1) Family member aged 65 or older; (2) University or high school student child; (3)  Junior high school student child; (4) Elementary school student child; (5) Infant or toddler child.

\paragraph{Q17. Occupation} \quad

\noindent\textbf{Question:} ``What is your occupation?''

\noindent\textbf{Response:} single choice ---
Office work /
Sales (shop owner, sales clerk, sales representative, etc.) /
Management (public servant or company employee at the section-chief level or higher; corporate officer, etc.) /
Professional or technical (teacher, engineer, legal professional, writer, artist, etc.) /
Medical or care work (doctor, nurse, certified care worker, etc.) /
Service (housekeeper, home helper, beautician/barber, hospitality, driver, security, etc.) /
Manual work (carpenter, repair worker, production-line worker, cleaner, etc.) /
Agriculture, forestry, or fishing /
Pension recipient /
Full-time homemaker /
Student /
Not working /
Other.

\paragraph{Q18. Household annual income (2022, including bonuses)} \quad

\noindent\textbf{Question:} ``What was your household's total taxable annual income for 2022, including bonuses?''

\noindent\textbf{Response:} single choice.
Categories:
under 1{,}000{,}000 /
1{,}000{,}000--under 2{,}000{,}000 /
2{,}000{,}000--under 4{,}000{,}000 /
4{,}000{,}000--under 6{,}000{,}000 /
6{,}000{,}000--under 8{,}000{,}000 /
8{,}000{,}000--under 10{,}000{,}000 /
10{,}000{,}000--under 12{,}000{,}000 /
12{,}000{,}000--under 14{,}000{,}000 /
14{,}000{,}000--under 16{,}000{,}000 /
16{,}000{,}000--under 18{,}000{,}000 /
18{,}000{,}000--under 20{,}000{,}000 /
20{,}000{,}000 JPY or more /
Do not know /
Prefer not to answer.

\paragraph{Q19. Personal annual income (2022, including bonuses)} \quad

\noindent\textbf{Question:} ``What was your personal taxable annual income for 2022, including bonuses?''

\noindent\textbf{Response:} single choice.
Categories:
No personal income /
under 1{,}000{,}000 /
1{,}000{,}000--under 2{,}000{,}000 /
2{,}000{,}000--under 4{,}000{,}000 /
4{,}000{,}000--under 6{,}000{,}000 /
6{,}000{,}000--under 8{,}000{,}000 /
8{,}000{,}000--under 10{,}000{,}000 /
10{,}000{,}000--under 12{,}000{,}000 /
12{,}000{,}000--under 14{,}000{,}000 /
14{,}000{,}000--under 16{,}000{,}000 /
16{,}000{,}000--under 18{,}000{,}000 /
18{,}000{,}000--under 20{,}000{,}000 /
20{,}000{,}000 JPY or more /
Do not know /
Prefer not to answer.

\newpage

\subsubsection{Selected Screenshots from Main Experiment}

\begin{figure}[htbp]
  \centering
  \includegraphics[width=0.8\linewidth]{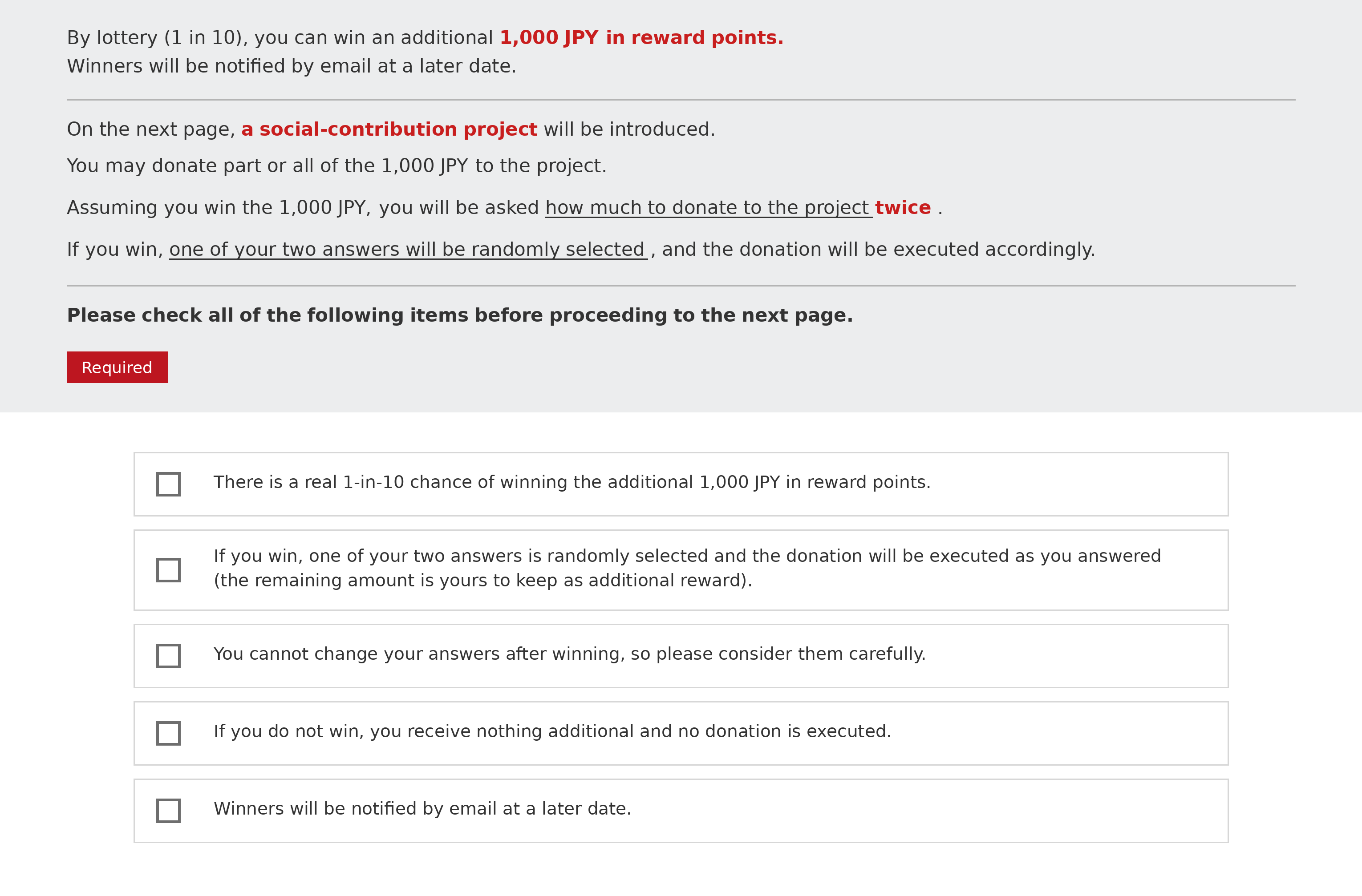}
  \caption{Q8SA: Lottery information and comprehension checklist.}
  \label{fig:q8sa_lottery_check}
\end{figure}

\begin{figure}[htbp]
  \centering
  \includegraphics[width=0.8\linewidth]{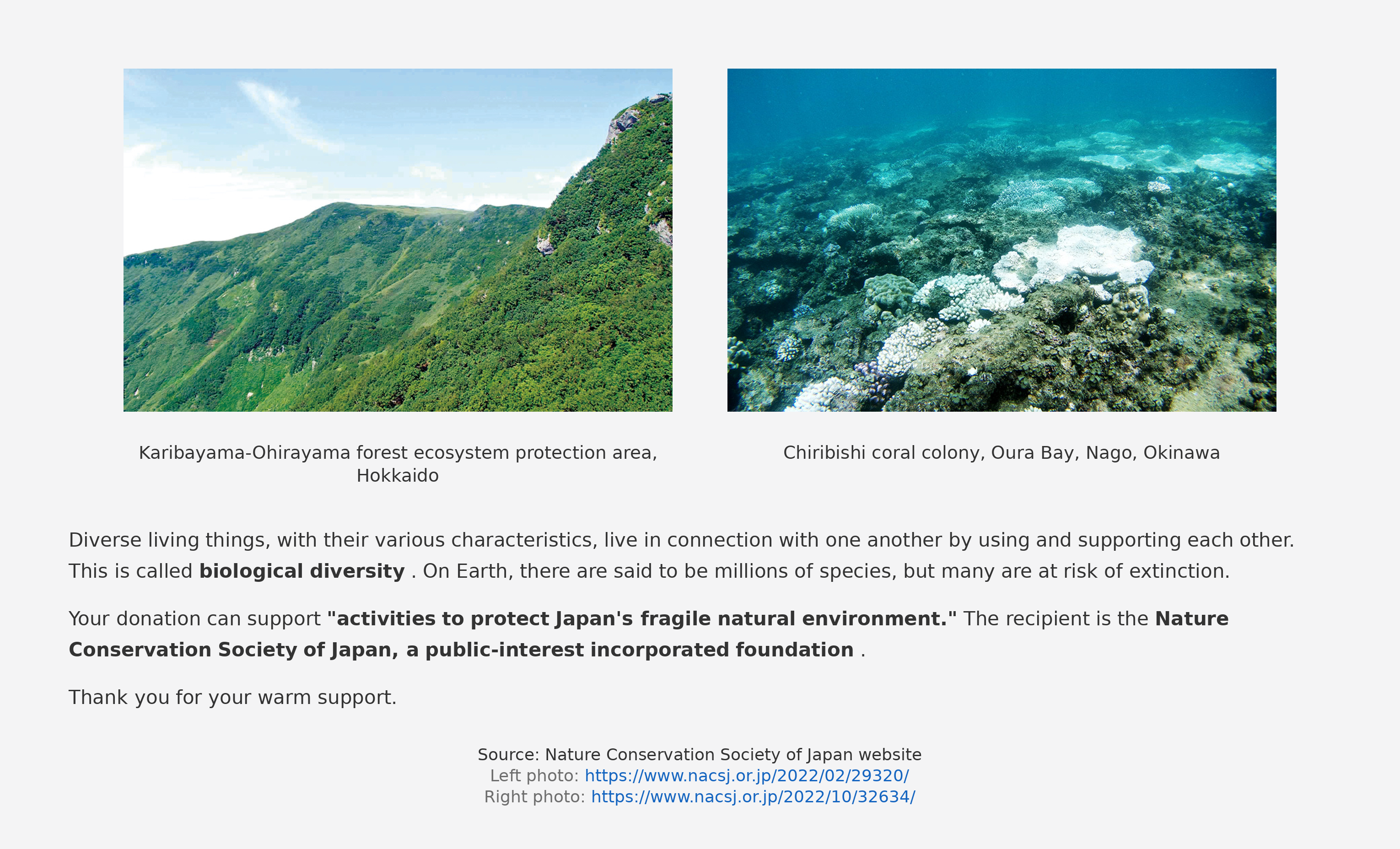}
  \caption{I8S: Introduction of the recipient charity.}
  \label{fig:i8s_charity_intro}
\end{figure}

\begin{figure}[htbp]
  \centering
  \includegraphics[width=0.8\linewidth]{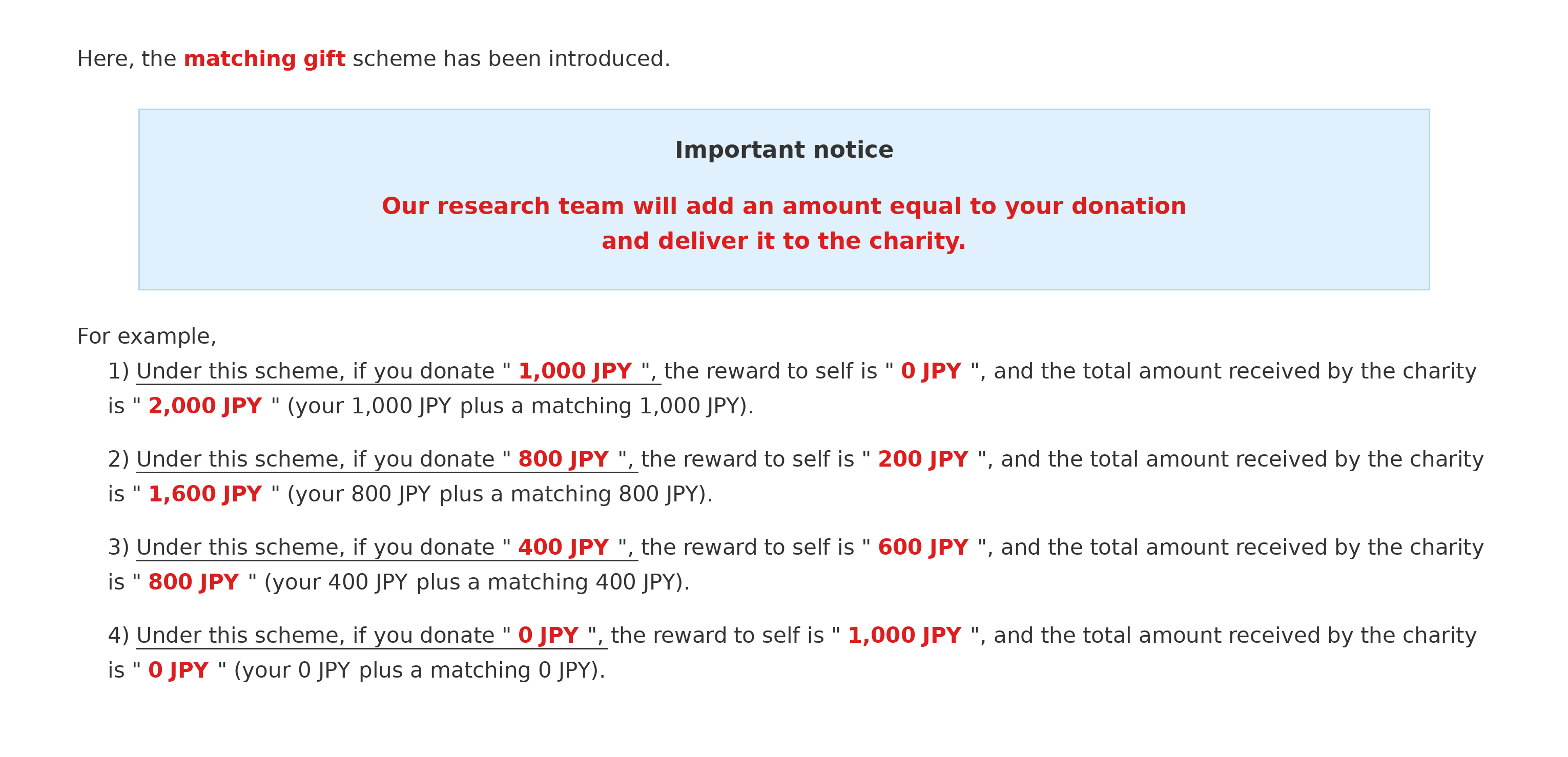}
  \caption{I8S2: Treatment instructions for the matching scheme (compulsory).}
  \label{fig:i8s2_match_compulsory}
\end{figure}

\begin{figure}[htbp]
  \centering
  \includegraphics[width=0.8\linewidth]{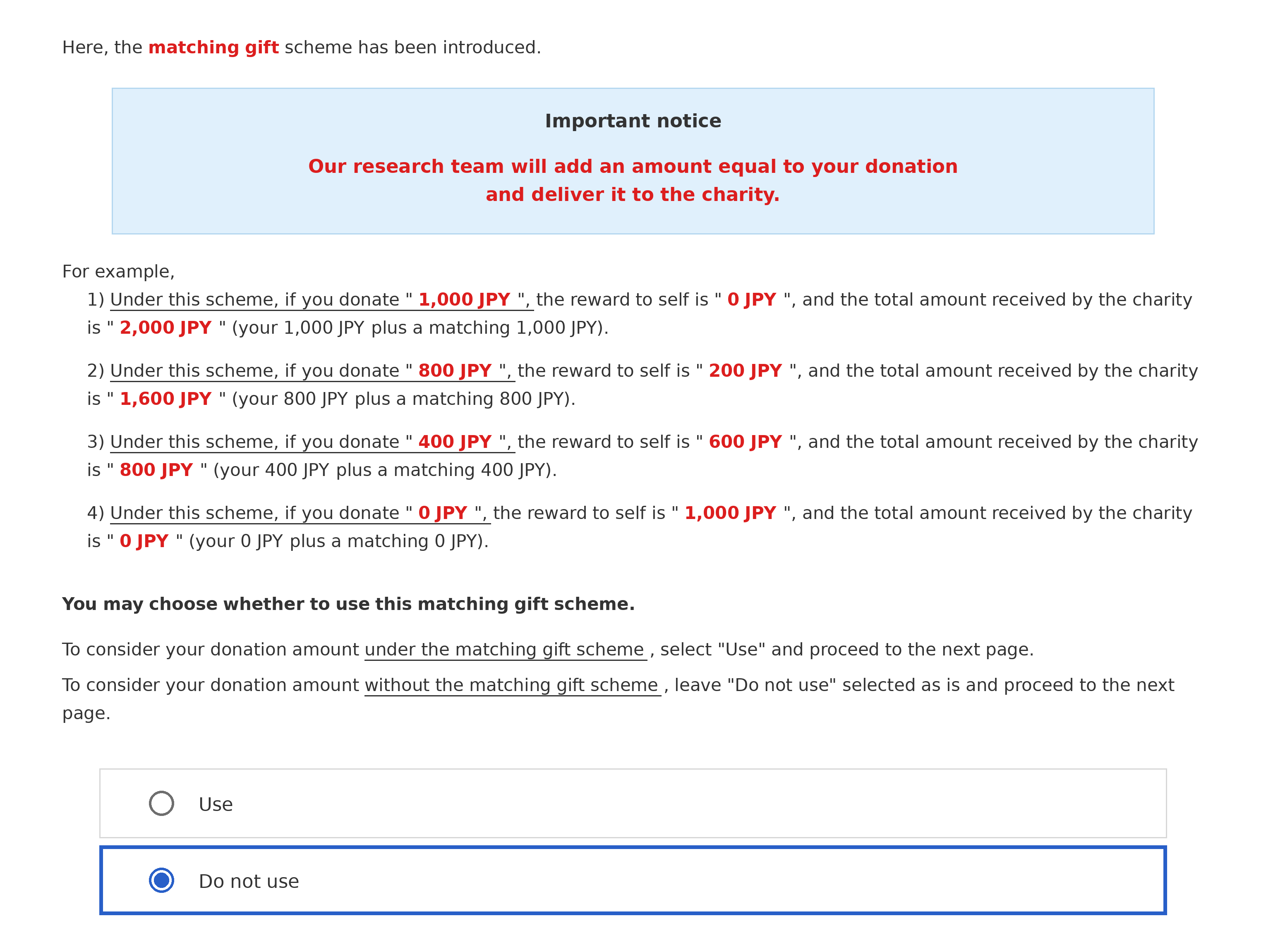}
  \caption{Q8S3O: Treatment instructions for the matching scheme (opt-in).}
  \label{fig:q8s3O_match_optin}
\end{figure}

\begin{figure}[htbp]
  \centering
  \includegraphics[width=0.8\linewidth]{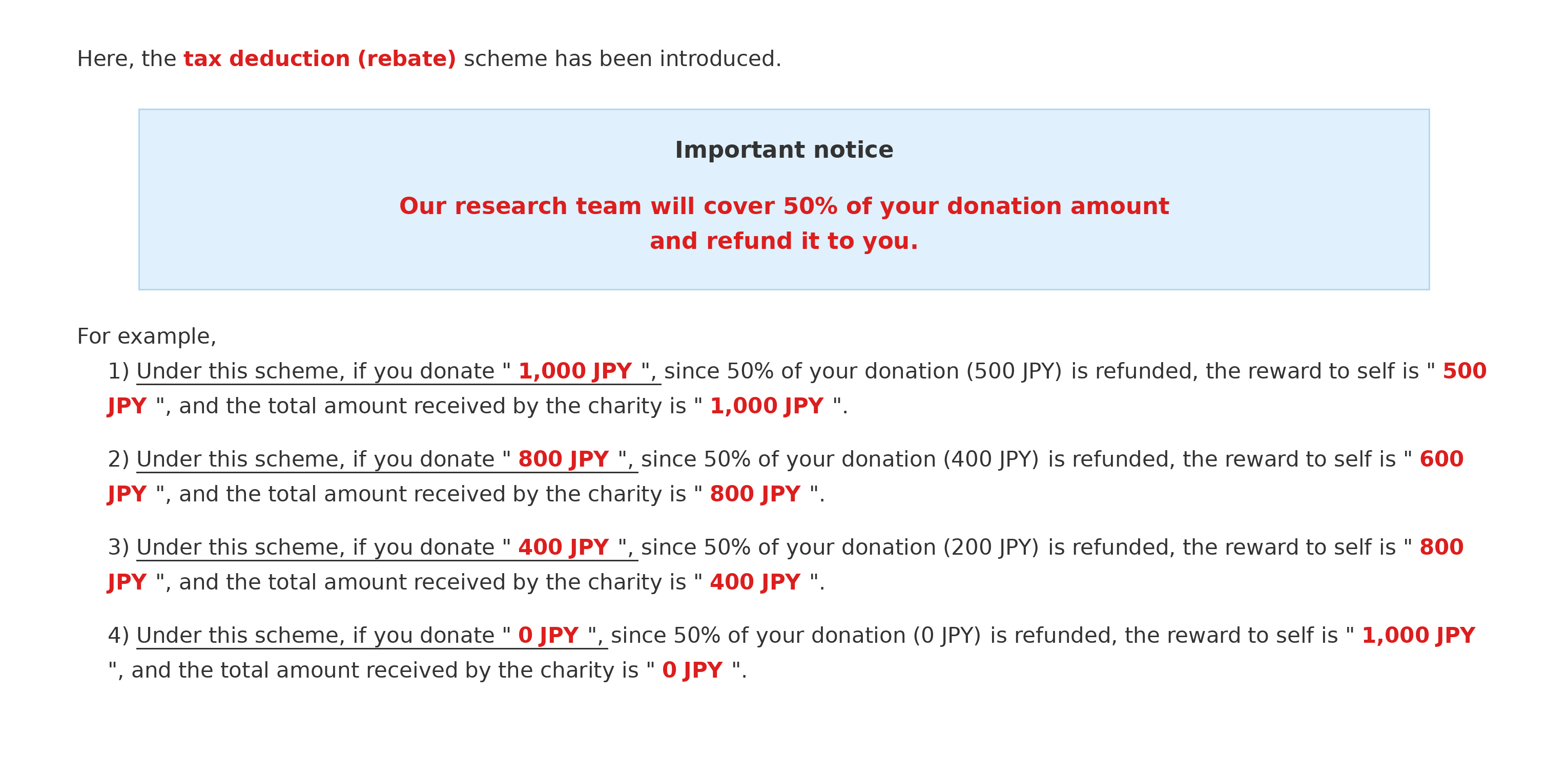}
  \caption{I8S4: Treatment instructions for the rebate scheme (compulsory).}
  \label{fig:i8s4_rebate_compulsory}
\end{figure}

\begin{figure}[htbp]
  \centering
  \includegraphics[width=0.8\linewidth]{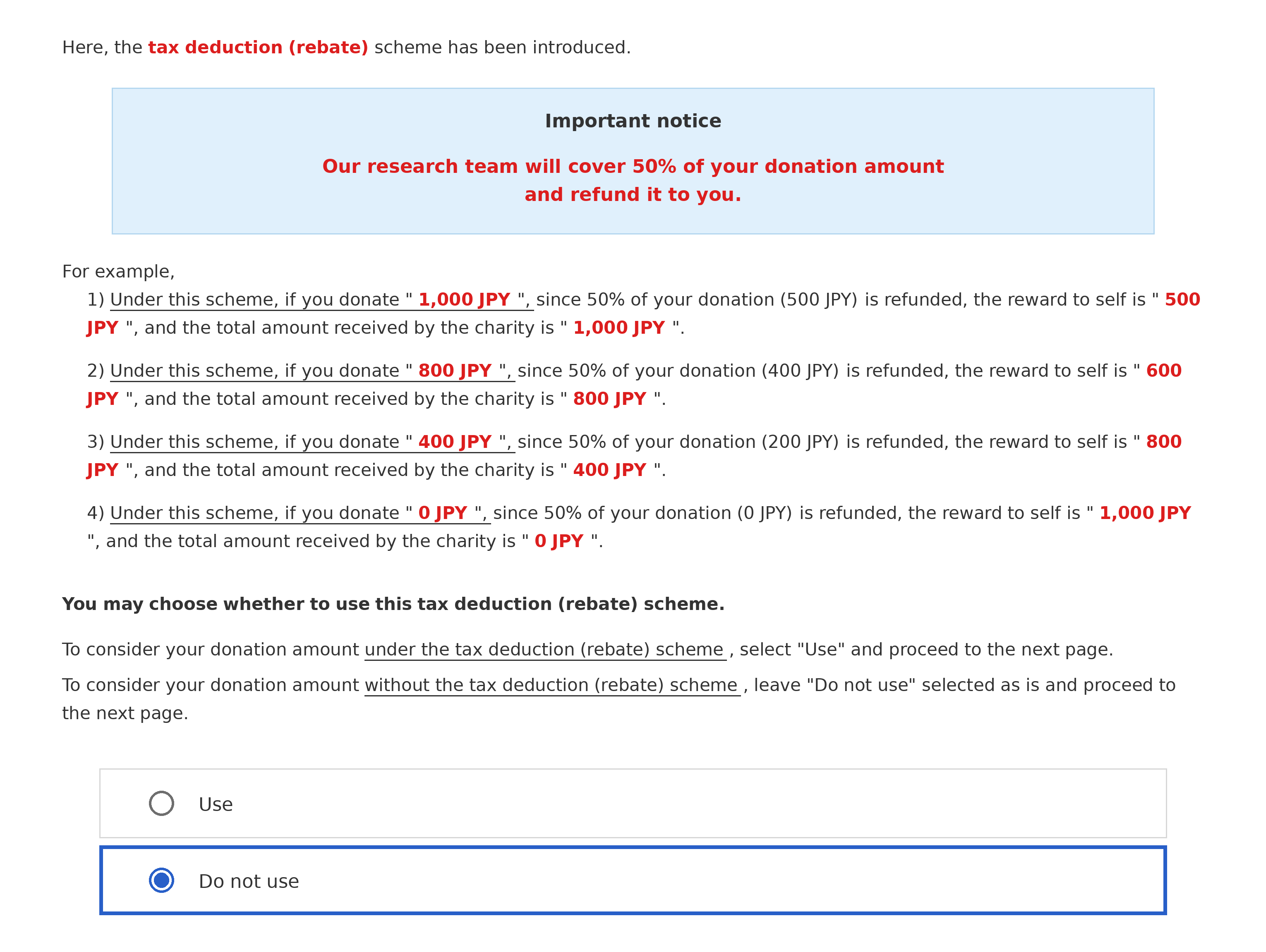}
  \caption{Q8S5O: Treatment instructions for the rebate scheme (opt-in).}
  \label{fig:q8s5O_rebate_optin}
\end{figure}

\begin{figure}[htbp]
  \centering
  \includegraphics[width=0.8\linewidth]{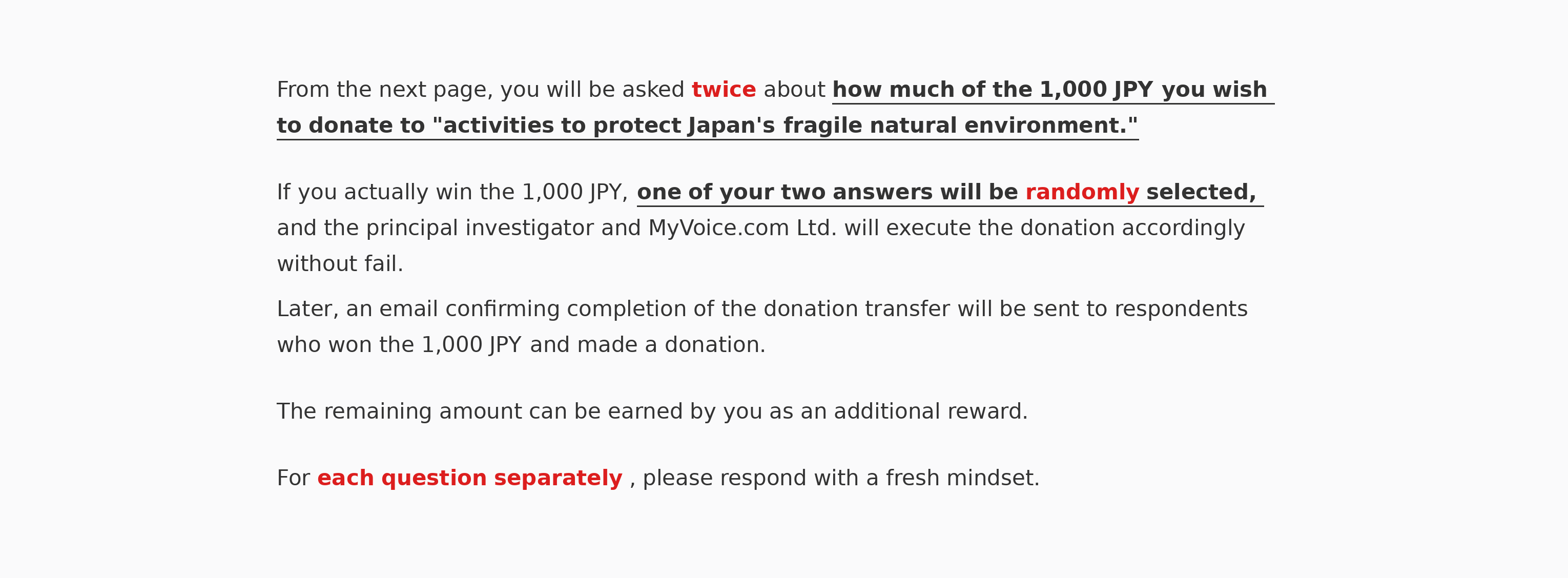}
  \caption{Pre-donation announcement.}
  \label{fig:predonation_notice}
\end{figure}

\begin{figure}[htbp]
  \centering
  \includegraphics[width=0.8\linewidth]{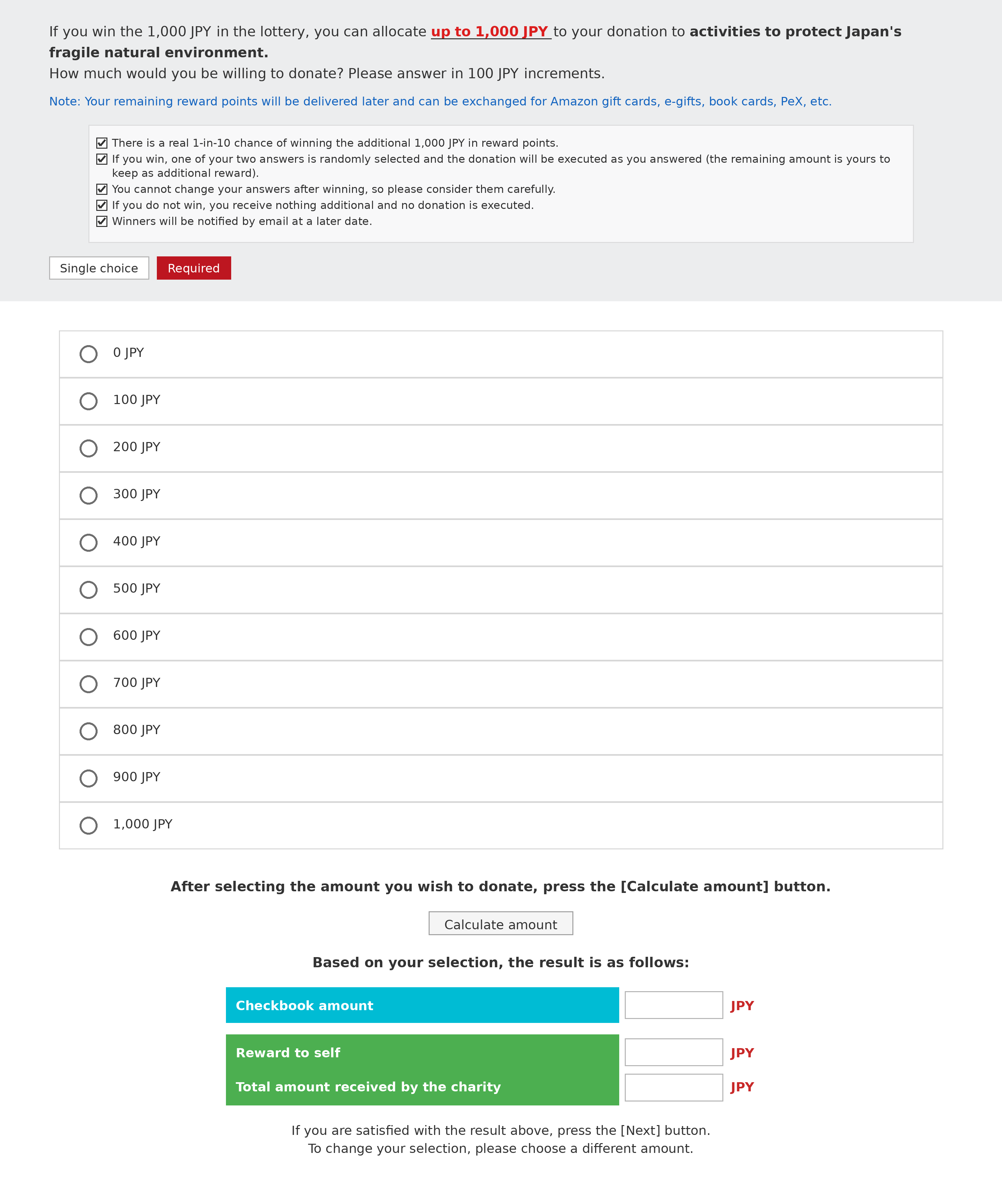}
  \caption{Q8S1B: Donation decision under the control condition (1,000 JPY cap).}
  \label{fig:q8s1b_control_1000}
\end{figure}

\begin{figure}[htbp]
  \centering
  \includegraphics[width=0.8\linewidth]{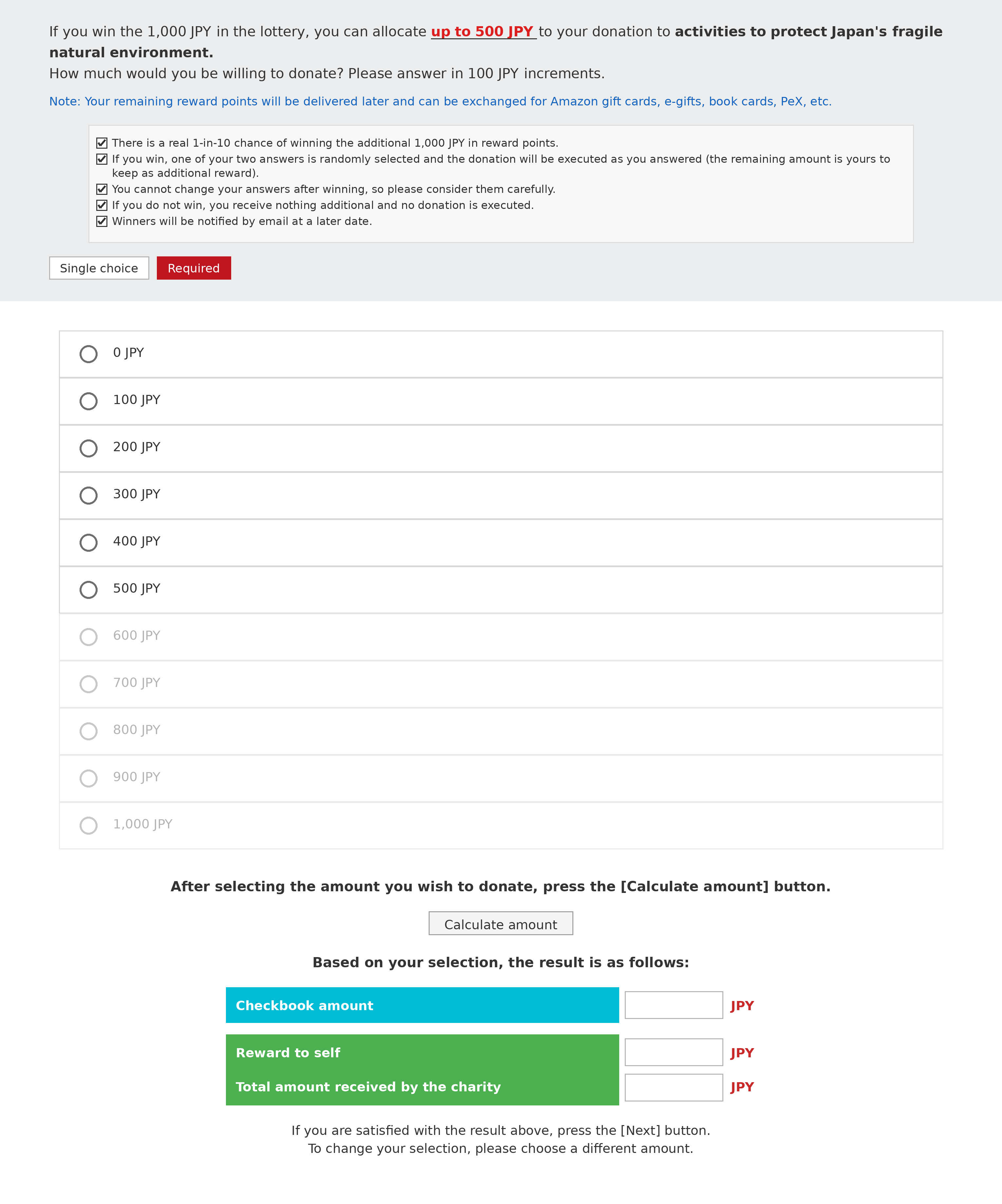}
  \caption{Q8S1C: Donation decision under the control condition (500 JPY cap).}
  \label{fig:q8s1c_control_500}
\end{figure}

\begin{figure}[htbp]
  \centering
  \includegraphics[width=0.8\linewidth]{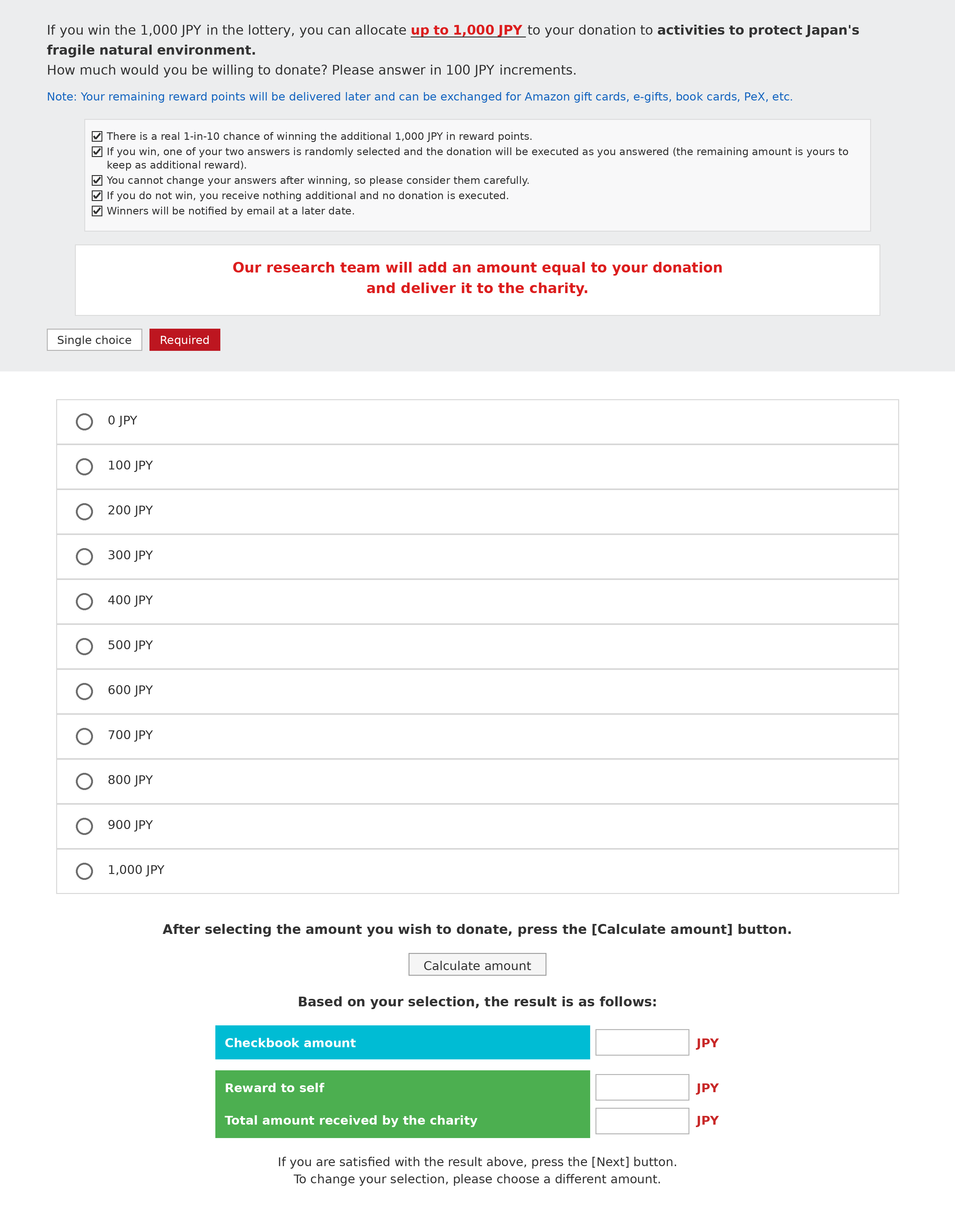}
  \caption{Q8S2B, Q8S3B: Donation decision under the matching gift scheme (1,000 JPY cap).}
  \label{fig:q8s2b3b_match_1000}
\end{figure}

\begin{figure}[htbp]
  \centering
  \includegraphics[width=0.8\linewidth]{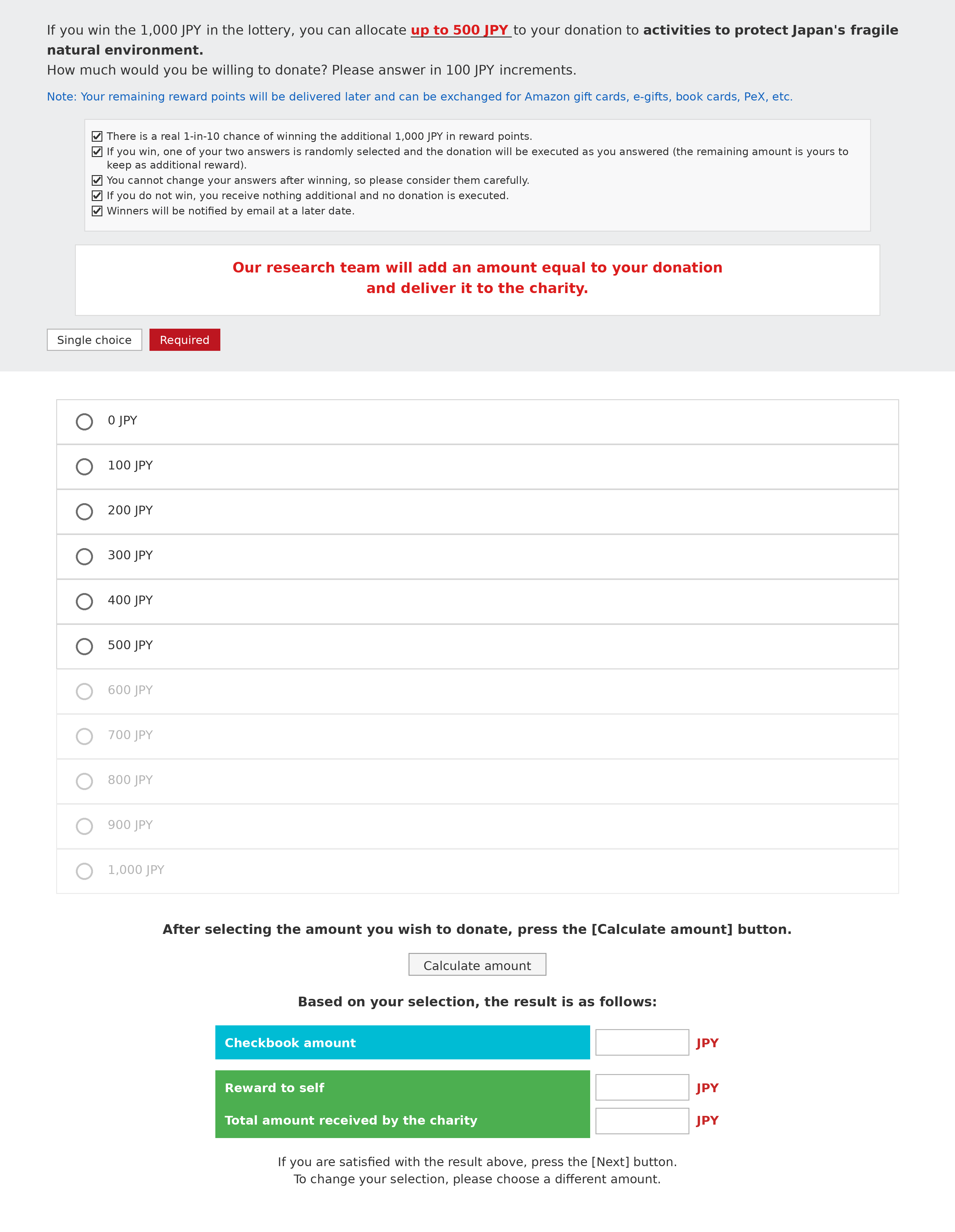}
  \caption{Q8S2C, Q8S3C: Donation decision under the matching gift scheme (500 JPY cap).}
  \label{fig:q8s2c3c_match_500}
\end{figure}

\begin{figure}[htbp]
  \centering
  \includegraphics[width=0.8\linewidth]{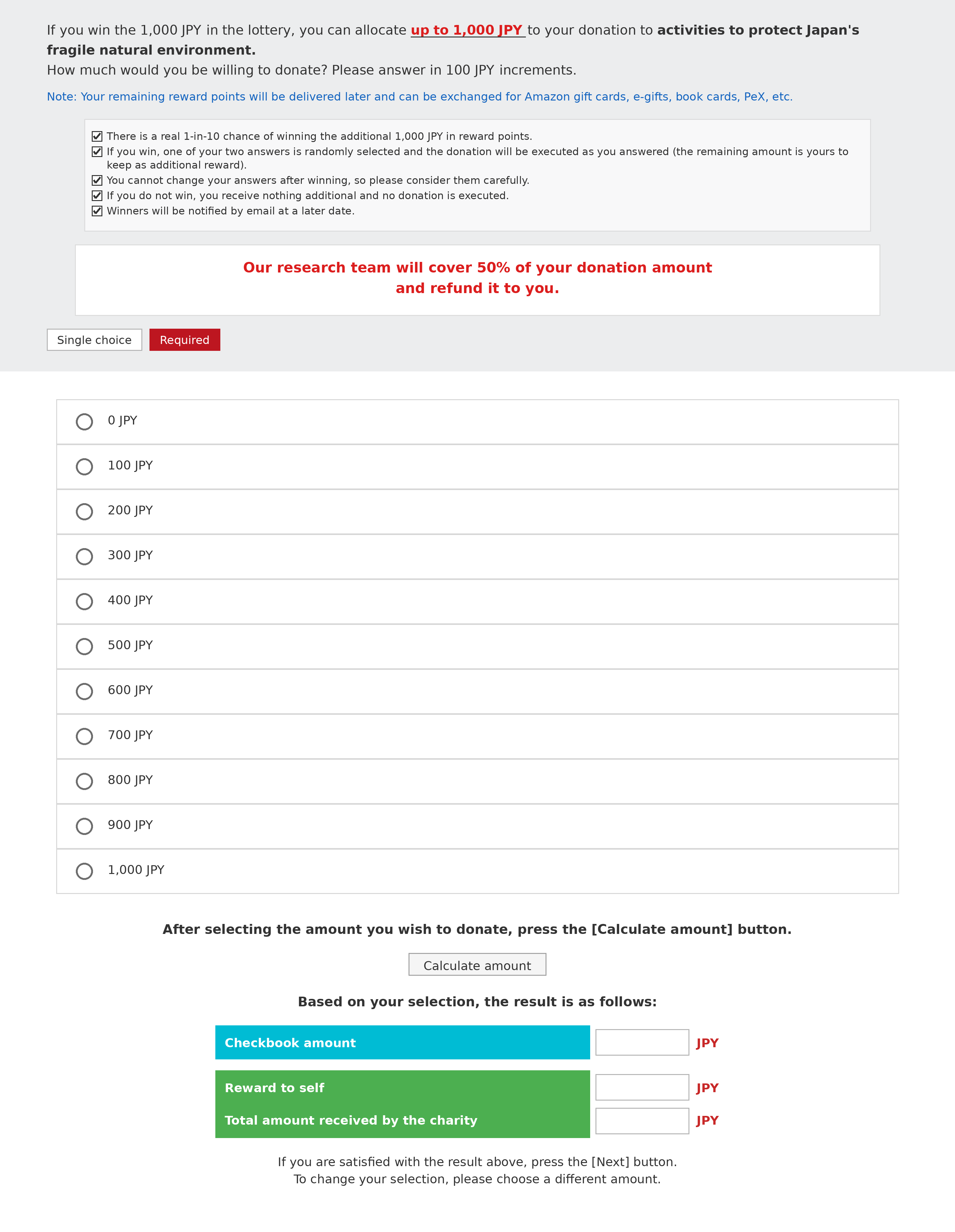}
  \caption{Q8S4B, Q8S5B: Donation decision under the rebate scheme (1,000 JPY cap).}
  \label{fig:q8s4b5b_rebate_1000}
\end{figure}

\begin{figure}[htbp]
  \centering
  \includegraphics[width=0.8\linewidth]{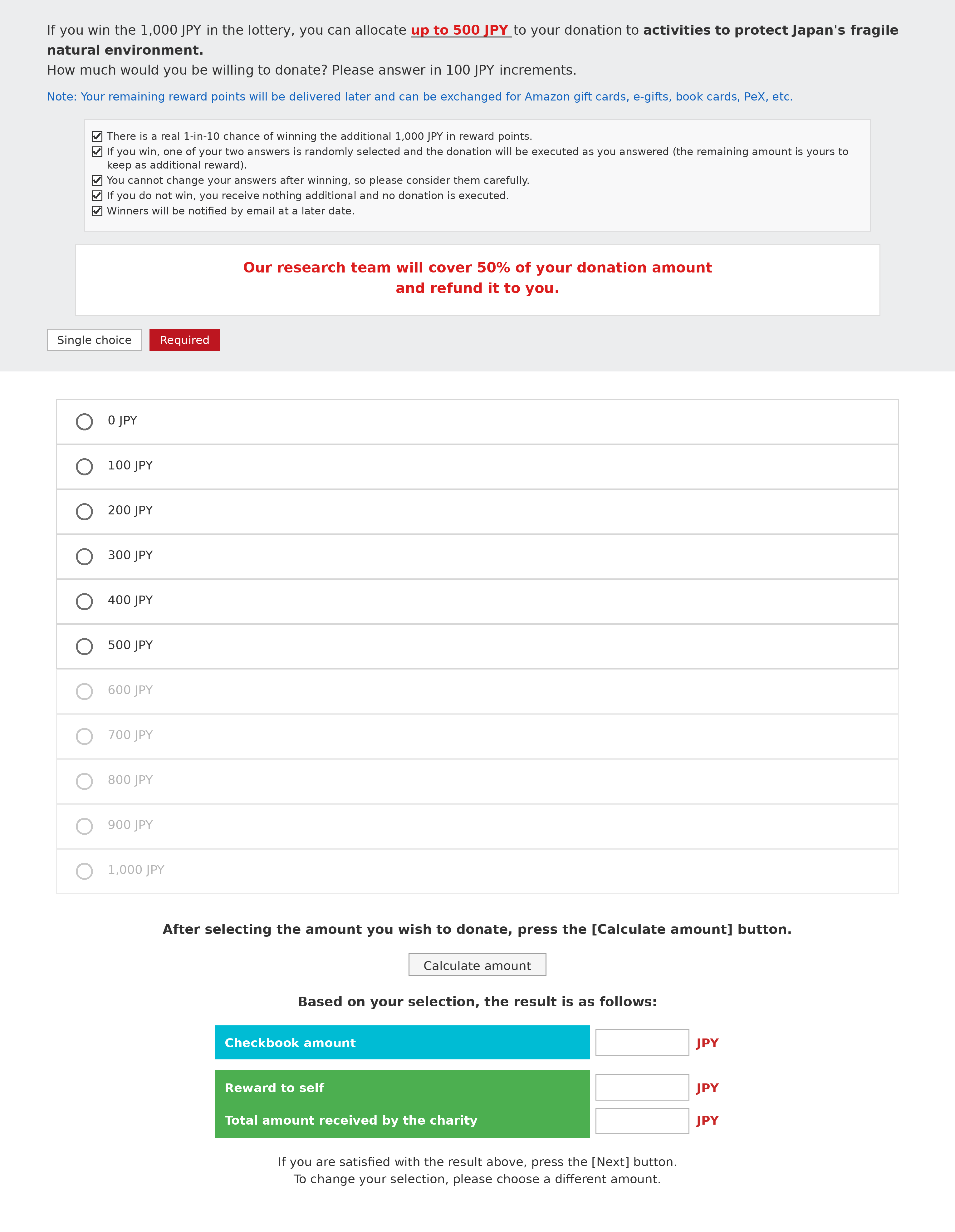}
  \caption{Q8S4C, Q8S5C: Donation decision under the rebate scheme (500 JPY cap).}
  \label{fig:q8s4c5c_rebate_500}
\end{figure}

\newpage

\setcounter{figure}{0}
\setcounter{table}{0}

\section{Additional Figures and Tables}
\label{app:additional}

\begin{figure}[htbp]
  \centering
  \includegraphics[width=0.8\linewidth]{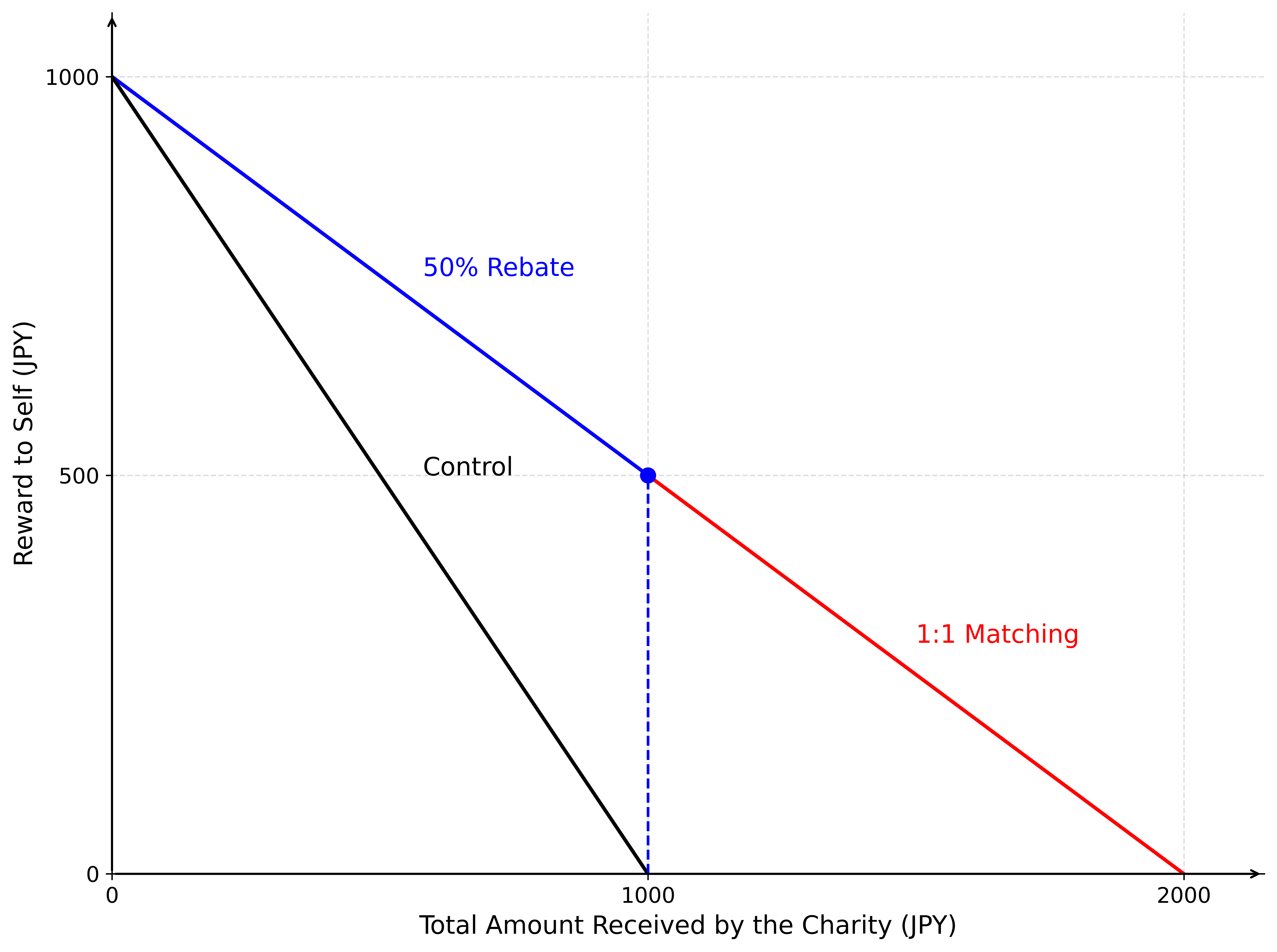}
  \caption{Budget Constraint Lines under Control, Rebate, and Matching.}
  \label{fig:budget_constraint}
\end{figure}

\newpage

\begin{sidewaystable}[htbp]
  \centering
  \caption{Example: Donation Outcomes under Each Treatment and Upper Limit}
  \label{tab:budget_example}
  \begin{threeparttable}
  \footnotesize
  \setlength{\tabcolsep}{6pt}
  \renewcommand{\arraystretch}{1.2}
  \begin{tabular}{l*{7}{c}}
    \toprule
     & Control & 1:1 matching & \multicolumn{2}{c}{1:1 matching} & 50\% rebate & \multicolumn{2}{c}{50\% rebate} \\
     &         & Compulsory   & \multicolumn{2}{c}{Opt-in}       & Compulsory  & \multicolumn{2}{c}{Opt-in}      \\
    \cmidrule(lr){4-5}\cmidrule(lr){7-8}
    \emph{Unit: JPY} & & & \cellcolor{gray!20}Take-up & No & & \cellcolor{gray!20}Take-up & No \\
    \midrule
    \textbf{Endowment}                    & \multicolumn{7}{c}{1{,}000} \\
    \textbf{Upper limit of checkbook amount}      & \multicolumn{7}{c}{1{,}000} \\
    \textbf{Upper limit of total amount received} & \cellcolor{gray!20}1{,}000 & 2{,}000 & 2{,}000 & \cellcolor{gray!20}1{,}000 & \cellcolor{gray!20}1{,}000 & \cellcolor{gray!20}1{,}000 & \cellcolor{gray!20}1{,}000 \\
    \hdashline
    \addlinespace[2pt]
    i. Checkbook amount                                & \cellcolor{gray!20}800 & 800     & 800     & \cellcolor{gray!20}800 & \cellcolor{gray!20}800 & \cellcolor{gray!20}800 & \cellcolor{gray!20}800 \\
    ii. Reward to self (Consumption)                   & \cellcolor{gray!20}200 & 200     & 200     & \cellcolor{gray!20}200 & \cellcolor{gray!20}600 & \cellcolor{gray!20}600 & \cellcolor{gray!20}200 \\
    \textbf{iii. Total amount received by the charity} & \cellcolor{gray!20}800 & 1{,}600 & 1{,}600 & \cellcolor{gray!20}800 & \cellcolor{gray!20}800 & \cellcolor{gray!20}800 & \cellcolor{gray!20}800 \\
    \midrule
    \textbf{Endowment}                    & \multicolumn{7}{c}{1{,}000} \\
    \textbf{Upper limit of checkbook amount}      & \multicolumn{7}{c}{500} \\
    \textbf{Upper limit of total amount received} & 500 & \cellcolor{gray!20}1{,}000 & \cellcolor{gray!20}1{,}000 & 500 & 500 & 500 & 500 \\
    \hdashline
    \addlinespace[2pt]
    i. Checkbook amount                                & 400 & \cellcolor{gray!20}400 & \cellcolor{gray!20}400 & 400 & 400 & 400 & 400 \\
    ii. Reward to self (Consumption)                   & 600 & \cellcolor{gray!20}600 & \cellcolor{gray!20}600 & 600 & 800 & 800 & 600 \\
    \textbf{iii. Total amount received by the charity} & 400 & \cellcolor{gray!20}800 & \cellcolor{gray!20}800 & 400 & 400 & 400 & 400 \\
    \bottomrule
  \end{tabular}
  \begin{tablenotes}
    \footnotesize
    \item \emph{Notes:} This table illustrates donation outcomes when each participant initially selects 80\% of the upper limit on the checkbook amount. Shaded cells indicate the values used in the main analysis. To equalize the maximum total amount received by the charity across treatments, the analysis uses the 1{,}000 JPY upper-limit responses for the control and rebate groups, and the 500 JPY upper-limit responses for the matching groups. In opt-in conditions, participants who did not take up the scheme (``No'' columns) face the same conditions as the control group; accordingly, the analysis uses the 1{,}000 JPY upper-limit responses for opt-in non-takers. At the time of the experiment, 1 USD was approximately 133 JPY.
  \end{tablenotes}
  \end{threeparttable}
\end{sidewaystable}

\newpage

\begin{table}[htbp]
  \centering
  \caption{Robustness Checks on the Treatment Effects under Self-selection}
  \label{tab:robustness_selfselection}
  \begin{threeparttable}
  \footnotesize
  \setlength{\tabcolsep}{3pt}
  \renewcommand{\arraystretch}{1.10}

  \begin{tabular}{@{}l*{4}{c}@{}}
    \toprule
    \emph{Treatment Effects}
      & (1) & (2) & (3) & (4) \\
    \addlinespace[2pt]

    Experimental design:
      & \multicolumn{4}{c}{With budget adjustment} \\
    \addlinespace[2pt]

    Endowment:
      & \multicolumn{3}{c}{1{,}000 JPY}
      & 10{,}000 JPY \\
    \addlinespace[2pt]

    Analysis sample:
      & \multicolumn{4}{c}{High-comprehension sample} \\
    \addlinespace[2pt]

    Model:
      & OLS
      & Tobit
      & LPM
      & OLS \\
    \addlinespace[2pt]

    Dependent variable:
      & \shortstack{Total giving\\amount}
      & \shortstack{Total giving\\amount}
      & \shortstack{Non-zero donation\\(1/0)}
      & \shortstack{Total giving\\amount} \\
    \midrule

    \textbf{1:1 matching}
      & 196.000***
      & 206.137***
      & 0.026
      & 1{,}469.339*** \\
    Opt-in (ITT)
      & (21.867)
      & (28.510)
      & (0.050)
      & (237.864) \\
    \addlinespace[4pt]

    \textbf{50\% rebate}
      & 101.846***
      & 107.422***
      & 0.030
      & 620.421** \\
    Opt-in (ITT)
      & (27.379)
      & (30.247)
      & (0.045)
      & (258.080) \\
    \addlinespace[8pt]

    \textbf{1:1 matching}
      & 242.435***
      & 221.939***
      & $-0.012$
      & 1{,}463.112*** \\
    Compulsory (ATE)
      & (25.264)
      & (31.415)
      & (0.047)
      & (205.438) \\
    \addlinespace[4pt]

    \textbf{50\% rebate}
      & 198.020***
      & 193.278***
      & 0.071
      & 1{,}245.293*** \\
    Compulsory (ATE)
      & (26.004)
      & (23.252)
      & (0.047)
      & (242.115) \\
    \addlinespace[8pt]

    Constant
      & $-50.116$
      & 420.042***
      & 0.835***
      & 1{,}766.518*** \\
      & (65.072)
      & (18.317)
      & (0.038)
      & (161.430) \\
    \midrule

    Matching (ITT) $-$ Rebate (ITT)
      & 94.154
      & 100.224
      & $-0.004$
      & 848.918 \\
    \quad $p$-value
      & 0.006
      & $<$0.001
      & 0.861
      & 0.005 \\
    \addlinespace[4pt]

    Matching (ATE) $-$ Rebate (ATE)
      & 44.415
      & 28.143
      & $-0.083$
      & 217.819 \\
    \quad $p$-value
      & 0.071
      & 0.295
      & 0.009
      & 0.443 \\
    \addlinespace[4pt]

    Take-up rate of 1:1 matching
      & 0.653 & 0.653 & 0.653 & 0.653 \\
    Take-up rate of 50\% rebate
      & 0.637 & 0.637 & 0.637 & 0.637 \\
    \addlinespace[4pt]

    Observations
      & 1{,}217 & 1{,}217 & 1{,}217 & 1{,}217 \\
    \addlinespace[2pt]

    $R^{2}$/Pseudo-$R^{2}$
      & 0.265 & 0.005 & 0.007 & 0.033 \\
    \bottomrule
  \end{tabular}

  \begin{tablenotes}
    \scriptsize
    \item \emph{Notes:} Cluster-robust standard errors at the regional level are reported in parentheses. *** $p<0.01$, ** $p<0.05$, * $p<0.1$. Column (1) reports OLS estimates with covariates comprising baseline altruism, age, years of education, household income, and indicators for being female, being married, having children, missing household-income information, and urban residence. Column (2) reports average marginal effects from a two-limit Tobit model accounting for censoring at both bounds. Its matching-minus-rebate contrasts and associated $p$-values are estimated directly using postestimation margins, rather than as arithmetic differences between the reported marginal effects; its ``Constant'' is the covariate-adjusted predicted outcome for the control condition, rather than the constant from the latent Tobit equation. Column (3) reports the extensive-margin analysis, and Column (4) reports the post-experiment hypothetical exercise with a tenfold-larger endowment of 10{,}000~JPY. The final row reports the pseudo-$R^{2}$ for Column (2) and the $R^{2}$ for the other columns.
  \end{tablenotes}
  \end{threeparttable}
\end{table}

\newpage

\begin{table}[htbp]
  \centering
  \caption{Multiple-Testing Adjusted $p$-values for Rebate-versus-Matching Contrasts}
  \label{tab:mht_corrections}
  \begin{threeparttable}
  \small
  \footnotesize
  \setlength{\tabcolsep}{10pt}
  \renewcommand{\arraystretch}{1.15}
  \begin{tabular}{l l c c c}
    \toprule
                          &                & Unadjusted   & Romano--Wolf & List--Shaikh--Xu \\
    Outcome               & Regime         & $p$-value    & $p$-value    & $p$-value   \\
                          &                & (regression) &              &             \\
    \midrule
    Checkbook giving      & Self-selection & $<0.001$ & $<0.001$ & $<0.001$ \\
    Checkbook giving      & Compulsory     & $<0.001$ & $<0.001$ & $<0.001$ \\
    Total amount received & Self-selection & 0.007    & 0.004    & 0.010    \\
    Total amount received & Compulsory     & 0.351    & 0.262    & 0.448    \\
    \bottomrule
  \end{tabular}
  \begin{tablenotes}
    \footnotesize
    \item \emph{Notes:} Unadjusted $p$-values are from cluster-robust Wald tests of the corresponding linear contrasts in the main specification (see Columns (1) and (2) of Table~3 in the main text); standard errors are clustered at the regional level. Romano--Wolf $p$-values are from \texttt{rwolf2} \citep{RomanoWolf2005} with 10{,}000 pairs-bootstrap replications, two-sided, with cluster-robust standard errors. List--Shaikh--Xu $p$-values are from Theorem~3.1 of \citet{ListShaikhXu2019}, computed via \texttt{mhtexp} with 10{,}000 bootstrap replications, one-sided; the package does not accommodate cluster-robust standard errors. The family consists of the four pre-registered primary-outcome contrasts of rebate versus matching across the self-selection and compulsory regimes, restricted to participants with high comprehension of the matching and rebate mechanisms (six or more correct out of eight on the comprehension check).
  \end{tablenotes}
  \end{threeparttable}
\end{table}

\newpage


\section{Pre-registration and Pre-analysis Plan}
\label{app:preregi}

\subsection{AEA RCT Registry Information}

\noindent
This study was pre-registered with the American Economic Association Randomized Controlled Trial Registry (AEA RCT Registry) prior to the randomized intervention and the collection of experimental outcomes. The registration details are as follows:
\begin{itemize}
    \item[-] \textbf{Title}: ``Rebate versus matching, again: Does opt-in matter?''
    \item[-] \textbf{AEA RCT Registry ID}: AEARCTR-0010943    
    \item[-] \textbf{DOI}: \texttt{https://doi.org/10.1257/rct.10943-1.1}    
    \item[-] \textbf{Initial registration}: February 13, 2023.
    \item[-] \textbf{Trial dates (as registered)}: The screening survey was conducted from February 10 to 15, 2023. The main experiment was scheduled to run from February 17 to 24, 2023, and was completed on February 21, 2023.
    \item[-] \textbf{Country}: Japan.
    \item[-] \textbf{Ethics approval}: This study received ex-ante approval from the IRB at The University of Osaka (Approval Number: 2022CRER0120-2, dated January 20, 2023).
\end{itemize}

\subsection{Summary of the Pre-analysis Plan}

\noindent
Alongside the AEA RCT Registry submission, the authors uploaded two supplementary documents that outlined the theoretical framework and the principal hypotheses prior to the randomized intervention and the collection of experimental outcomes:
``Theoretical Background'' (February 12, 2023; 5 pages) and
``Hypothesis Setting and Analysis Plan'' (February 12, 2023; 5 pages).
The following key elements were pre-specified.

\paragraph{Theoretical framework}
The impure impact giving model of \citet{hungerman2021impure} was adopted as the underlying theoretical model. Optimal levels of \textit{checkbook giving} $g^*$ and \textit{total amount received} $G^*$ were derived under the rebate price $p_t$ and matching price $p_m$, and Slutsky decompositions yielded the rebate-price elasticity ($e_t$), the matching-price elasticity on the total amount received ($e_m$), and the matching-price elasticity on checkbook giving ($e_{m,o}$).\footnote{In the notation of Appendix~A.1, $e_t = e^g_r = e^G_r = e$, $e_m = e^G_m$, and $e_{m,o} = e^g_m$.}

\paragraph{Pre-registered predictions}
Three predictions were specified:
\begin{enumerate}
    \item On the total amount received by the charity:     
    if warm-glow preferences weaken the conventional substitution effect, $e_m > e_t$;
    if they strengthen it, $e_t > e_m$; 
    and if there are no warm-glow preferences, $e_t = e_m$.
    
    \item On checkbook giving:     
    $e_{m,o} > e_t$ for any preference. 
    That is, the matching incentive is less effective than the rebate incentive in increasing the donor’s checkbook giving.
    
    \item On the opt-in decision:     
    (a) when matching and rebate yield identical donation prices, the rebate is weakly preferred to matching for any donor preference (with equality for pure-impact donors, and no opt-in benefit from matching for pure warm-glow donors);
    (b) donors with larger treatment-induced increases in donations are more likely to take up the scheme.
\end{enumerate}

\paragraph{Experimental design and outcomes}
Five experimental arms were registered:
Control,
Compulsory Matching (CM),
Self-selection Matching (SM),
Compulsory Rebate (CR),
and Self-selection Rebate (SR).
Three primary outcomes were pre-registered:
(i) the \textit{initially selected amount} (the donor’s choice before any matching or rebate adjustment),
(ii) the \textit{actual donation expenditure} (the donor’s net out-of-pocket cost after accounting for any rebate refund),
(iii) the \textit{total amount donated to the charity} (the amount delivered to the recipient, including matched funds under matching).
The take-up (opt-in) probability was registered as a secondary outcome.
Stratified randomization by age, sex, and residential area, a sample size of 2{,}400 with 480 per arm, and the corresponding power calculation are documented in the AEA RCT Registry entry (see D.1).

\paragraph{Pre-specified analyses}

The plan specified linear ITT regressions of each primary outcome on the four treatment-group dummies, and an instrumental-variable (IV) regression to identify the Treatment-on-the-Treated (TOT) effect among self-selectors in the SM and SR groups using random assignment as an instrument.
The plan also envisaged estimation of the Treatment-on-the-Untreated (TOU) effect among self-selection non-takers, via comparison with the compulsory groups, following the approach of \citet{fowlie2021default}.
The following null hypotheses were stated for formal testing:

\[
\hat{\tau}^{ITT}_{CM} = \hat{\tau}^{ITT}_{CR},
\qquad
\hat{\gamma}_{SM} = \hat{\gamma}_{SR}.
\]

\[
\hat{\tau}^{ITT}_{CM} = \hat{\tau}^{TOT}_{SM},
\qquad
\hat{\tau}^{ITT}_{CR} = \hat{\tau}^{TOT}_{SR},
\qquad
\hat{\tau}^{TOT}_{SM} = \hat{\tau}^{TOT}_{SR}.
\]

\paragraph{Design adjustment for budget constraints}

The plan also specified an adjustment to equalize the maximum total amount received by the charity across treatment comparisons by varying the upper limit on the checkbook amount choice. In the budget-adjusted analysis, we use the 1,000 JPY upper-limit responses for the control and rebate conditions and the 500 JPY upper-limit responses for the matching conditions; in the opt-in conditions, non-takers are analyzed using the corresponding control-condition responses (see Section~3.4 of the main text and Appendix Table~\ref{tab:budget_example}).

\paragraph{Mapping to the hypotheses in the main text}

The four hypotheses H1--H4 derived in Appendix A.1--A.4 correspond to the pre-registered predictions as follows:

\begin{itemize}
    \item[-] \textbf{H1} (rebate $>$ matching on checkbook giving, regardless of warm-glow) reflects Theoretical Background Prediction 2.
    \item[-] \textbf{H2} (rebate vs.\ matching on the total amount received, with direction depending on $\gamma$ and $e$) reflects Prediction 1.
    \item[-] \textbf{H3} (take-up under self-selection: rebate $\geq$ matching when $\gamma > 0$) reflects Prediction 3, part (a).    
    \item[-] \textbf{H4} (TOT $>$ ATE under self-selection) reflects Prediction 3, part (b).
\end{itemize}
The substantive content of the pre-registered predictions is preserved; the main text refines their formal statement using the signed elasticity convention and explicit warm-glow conditions. In the pre-analysis-plan notation, $\hat{\gamma}_{SM}$ and $\hat{\gamma}_{SR}$ denote take-up rates rather than the warm-glow parameter $\gamma$, and the compulsory-condition ITT effects correspond to the ATEs reported in the main text.

\subsection{Participant Recruitment and Sample Construction}
\label{app:sample}

Participants were recruited through MyVoice Communications, a Japanese survey company that operates a large online research panel. The screening survey was completed by 10,022 respondents. Recruitment for the screening survey used quota cells based on Japan's 2020 Population Census, the official national census, defined by sex, 10-year age group, and region, yielding 100 cells in total.

Among the screening-survey completers, 6,463 respondents were invited to the main experiment. The remaining respondents were not excluded on substantive or data-quality grounds. Instead, invitations were sent sequentially as part of the survey company's standard fielding procedure, and the fielding was closed once the target number of completed responses had been reached.

The main experiment was completed by 2,400 respondents. For each of the five experimental groups, completed responses were collected using the same 100 quota cells based on sex, 10-year age group, and region. Within each cell, respondents were randomly assigned to one of the five experimental groups through the survey company's automated allocation procedure.

Respondents who triggered the attention-check prompt were retained in the main analysis sample. The prompt was shown immediately after a response that did not follow the instruction embedded in the preceding item, and asked respondents to read each question carefully. This procedure helps ensure that subsequent responses were provided after the reminder, thereby mitigating concerns about unnoticed inattention. The analysis sample therefore includes these respondents unless otherwise specified.

\subsection{Extensions to the Pre-analysis Plan}

\noindent
The analyses reported in the main text follow the pre-analysis plan in their core structure (ITT, IV/TOT, and TOU estimation on the pre-registered primary outcomes), but extend the plan in the following respects. We disclose these extensions for transparency.

\begin{enumerate}
    \item \textbf{Structural parameter estimation.}
    Section 4.2 estimates the underlying price elasticity of giving ($e$) and the warm-glow preference parameter ($\gamma$) using the Poisson regression approach of \citet{chen2024logs}, which was first circulated after our pre-registration.
    This structural estimation was therefore not pre-specified in the pre-analysis plan, which envisaged only reduced-form ITT and IV/TOT regressions; we report it as a complement to the reduced-form analyses.
    
    \item \textbf{Focal-sample restriction by comprehension.}    
    The main reduced-form analysis in Section 4.1 focuses on participants with a high level of understanding of the matching and rebate mechanisms (six or more correct answers out of eight on the comprehension check).

    This focal-sample choice follows established practice in the matching-versus-rebate literature, where restricting attention to (or otherwise accounting for) participants who comprehend the schemes is a standard step. Building on the seminal comparison of \citet{eckel2003rebate}, the subsequent ``isolation effect'' literature (\citealt{davis2005rebates,davis2005subsidy,davis2006rebate}) documents that misunderstanding of the mechanisms can contribute to the matching-versus-rebate gap.

    Although the pre-analysis plan did not pre-specify this restriction, we adopt it to align our analysis with this literature and to isolate responses that are not confounded by misunderstanding. To keep its contribution fully transparent, Section 4.1 reports the results in a stepwise fashion, beginning with the full pre-registered sample and progressively narrowing to the comprehension-restricted sample, so that the effect of the restriction is visible at every step.
    
    \item \textbf{Exploratory analysis of selection mechanisms.}    
    Section 5.3 reports two exploratory analyses not pre-specified in the plan:
    (a) a regression of opt-in take-up on baseline altruism and other covariates, and
    (b) Treatment-on-the-Treated effects on the donor's own reward (consumption).
    These analyses were motivated by the asymmetric selection pattern observed in the pre-registered LATE estimates.
    
    \item \textbf{Outcome labels and selection.}    
    The pre-analysis plan registered three primary outcomes.
    The main text reports the pre-registered \textit{``initially selected amount''} under the equivalent label \textit{``checkbook giving''} and the pre-registered \textit{``total amount donated to the charity''} under the label \textit{``the total amount received by the charity''}; both relabelings follow established usage in the recent rebate-versus-matching literature.    
    The remaining pre-registered outcome, the \textit{``actual donation expenditure,''} is deliberately omitted from the main analyses because it is a linear transformation of the total amount received, so reporting it separately would be largely redundant.
\end{enumerate}

\end{document}